\documentclass[aps,pra,reprint,amssymb,superscriptaddress]{revtex4-1} 
\usepackage{amsmath,amssymb,amsfonts}
\usepackage{graphicx}
\usepackage{dcolumn}
\usepackage{bm}
\usepackage[colorlinks=true,citecolor=blue,linkcolor=blue,urlcolor=blue]{hyperref}
\usepackage[mathlines]{lineno}

\usepackage{float}
\makeatletter
\let\newfloat\newfloat@ltx
\makeatother
\usepackage{algorithm}
\usepackage{algpseudocode}

\usepackage{breqn}
\usepackage{braket}
\usepackage[dvipsnames]{xcolor}
\usepackage{tabularray}

\begin{document}

\title{Variational protocols for emulating digital gates\\ using analog control with always-on interactions}

\author{Claire Chevallier}
\let\comma,
\affiliation{PASQAL SAS\comma{} 7 rue Léonard de Vinci - 91300 Massy\comma{} Paris\comma{} France}

\author{Joseph Vovrosh}
\email{joseph.vovrosh@pasqal.com}
\let\comma,
\affiliation{PASQAL SAS\comma{} 7 rue Léonard de Vinci - 91300 Massy\comma{} Paris\comma{} France}

\author{Julius de Hond}
\let\comma,
\affiliation{PASQAL SAS\comma{} 7 rue Léonard de Vinci - 91300 Massy\comma{} Paris\comma{} France}

\author{Mario Dagrada}
\let\comma,
\affiliation{PASQAL SAS\comma{} 7 rue Léonard de Vinci - 91300 Massy\comma{} Paris\comma{} France}

\author{Alexandre Dauphin}
\let\comma,
\affiliation{PASQAL SAS\comma{} 7 rue Léonard de Vinci - 91300 Massy\comma{} Paris\comma{} France}

\author{Vincent E. Elfving}
\let\comma,
\affiliation{PASQAL SAS\comma{} 7 rue Léonard de Vinci - 91300 Massy\comma{} Paris\comma{} France}

\begin{abstract}
    We design variational pulse sequences tailored for neutral atom quantum simulators and show that we can engineer layers of single-qubit and multi-qubit gates. As an application, we discuss how the proposed method can be used to perform refocusing algorithms, SWAP networks, and ultimately quantum chemistry simulations. While the theoretical protocol we develop still has experimental limitations, it paves the way, with some further optimisation, for the use of analog quantum processors for variational quantum algorithms, including those not previously considered compatible with analog mode.
\end{abstract}

\date{\today}

\maketitle

\section{\label{sec:intro}Introduction}

Quantum computers present an exciting alternative approach to their classical analogues. By leveraging special-purpose compute elements, which exhibit quantum properties, such as qubits, it has the potential to give exponential speed ups to very specific problems such as the simulation of quantum systems~\cite{doi:10.1126/science.273.5278.1073} and breaking cryptographic codes~\cite{365700}. The most common approach to quantum computing is digital, in which a set of one-qubit and two-qubits gates are used to build a theoretically universal quantum computer. There have been attempts to build such a universal device with various technologies, for example, superconducting circuits~\cite{PhysRevA.76.042319}, trapped ions~\cite{kielpinski2002architecture, bruzewicz2019trapped}, neutral atoms~\cite{henriet2020quantum, Bluvstein_2023} and photonic devices~\cite{slussarenko2019photonic}.  

Currently, a fully fault-tolerant quantum computer remains out of reach but the last decade has seen much progress towards this goal. Quantum error correction protocols have been proposed to remove these errors but, before they can be implemented at a meaningful scale, the number of qubits in Noisy Intermediate-Scale Quantum (NISQ) devices and the noise level must be improved by many orders of magnitude~\cite{PhysRevX.11.041058}.  As the quantum circuits that are promised to solve real world problems require large numbers of qubits and deep circuits, for example due to trotterisation~\cite{trotter}, these devices are not yet practically useful. Despite these issues, there have been impressive results in recent years, for example, solving classically intractable sampling problems~\cite{arute2019quantum, doi:10.1126/science.abe8770}, and accurate quantum simulation of a spin system using more than 100 qubits~\cite{kim2023evidence}, albeit for a classically tractable problem~\cite{patra2023efficient, tindall2023efficient, liao2023simulation}. As a result of these successes, quantum computers have gathered much interest from researchers in fields such as quantum chemistry~\cite{elfving2020quantum,daley2022practical}, the financial sector~\cite{orus2019quantum}, pharmaceuticals~\cite{cao2018potential}, and artificial intelligence~\cite{schuld2021machine, Bharti_2022, Cerezo_2021, dawid2023modern}.

\begin{figure}
    \centering
    \includegraphics{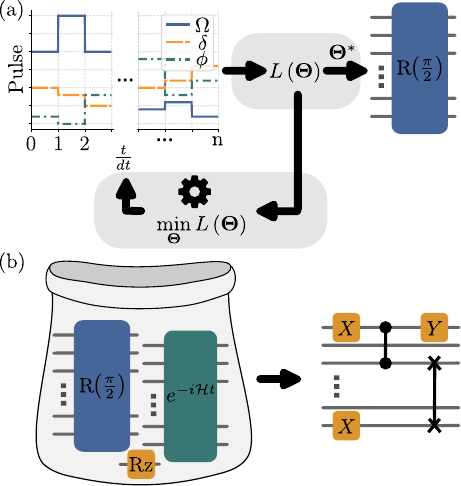}
    \caption{A schematic showing how to use an analog pulse sequence to approximate a variety of one- and two-qubit gates on neutral atom hardware. (a) We variationally prepare pulse sequences to achieve global $\pi/2$ rotations through tuning \textit{global} Rabi frequencies, detunings and phases of external lasers. (b) These global rotations along side single-qubit Rz gates and analog pulses of Eq.~\eqref{eq_ising} lead to a general set of gates available in a \textit{purely} analog fashion.}
    \label{fig_intro}
\end{figure}

 Analog quantum processors are alternative avenue for solving complex problems. By carefully isolating a given register of particles and understanding the underlying physics that governs the interactions between these particles we have access to and control over a resource Hamiltonian which we can use to encode and solve problems of interest. In an attempt to build analog devices, technologies such as trapped ions~\cite{Zhang_2017}, cold atoms~\cite{Choi_2016} and photonics~\cite{Tillmann_2013} have been used. However, in this work we focus our attention on neutral atoms as the particles in use~\cite{henriet2020quantum, briegel2000quantum}. While analog devices are not universal, and are thus more limited in their potential use cases, they have lower error rates than current digital quantum computers and therefore are likely to show the first industry-relevant quantum advantage~\cite{bloch2012quantum, Preskill_2018}.

Naturally, as analog devices are closely related to quantum experimental setups, applications for quantum simulation are the most promising contender for early quantum advantage. In fact, there have been experimental results published that challenge state-of-the-art classical simulation methods already~\cite{Choi_2016, Scholl_2021, shaw2023benchmarking}. Generalising the use of an analog system to problems outside the realm of quantum simulation is tough. One issue is that of \textit{always-on} interactions; typically in physical systems, interactions between particles scale with their separation. While the exact scaling is defined by the type of particles in use the consequence is the same, for a fixed register of atoms the interactions are fixed. As a result, performing quantum operations that act on an isolated part of the quantum system is difficult. This greatly limits the potential applications for analog quantum devices. 

A third regime of interest is Digital-Analog Quantum Computing (DAQC). Here, a quantum device has access to both analog pulses along-side single-qubit gates. NMR setups are known to work within this regime~\cite{JONES201191}, and proposals have considered realising a DAQC device in superconducting qubits~\cite{Lamata_2018, PRXQuantum.2.020328, PhysRevResearch.2.033103} with some early stage experimental results~\cite{greenaway2023analogue}. In fact, under the right conditions, this type of quantum computing can be universal~\cite{parra2020digital, garciadeandoin2023digitalanalog}. There are many proposed applications for such a device, for example, quantum chemistry~\cite{ kumar2023digitalanalog}, implementing the quantum approximate optimization and  Harrow-Hassidim-Lloyd algorithms~\cite{Headley_2022,PhysRevApplied.19.064056}, and computing the quantum Fourier transformation~\cite{Martin_2020}. However, in order to realise single-qubit gates previous work has assumed either that interactions can be turned off, external laser strengths much larger than these interactions - regimes that are difficult to realise for many technologies, or the availability of analog pulses with lengths that are below experimental feasibility.

There are many works that consider quantum devices at pulse level and propose methods to reach either the DAQC or digital regime, for example in superconducting hardware~\cite{PhysRevResearch.5.033159, PhysRevA.109.012616} and neutral atoms devices~\cite{PhysRevLett.131.170601}. Given a simple \textit{black box} of such pulse sequences, generic quantum circuits can be realised through compilation methods. A common choice is the refocusing algorithm~\cite{JONES1999322, PhysRevA.61.042310, PhysRevA.65.040301, Guseynov_2022}. However, a promising method recently developed considers the use of Walsh sequences, see Ref~\cite{votto2024robust}. We aim to develop such a compilation method with a neutral atom device in mind, with minimal experimental requirements.

In this work, we show how to realise a general set of single-qubit gates for a device that has always-on interactions in the regime where the external laser fields and pulse lengths are \textit{on the same order} as the interactions between qubits. In particular, we show how a surprisingly simple variational protocol requiring just access to single-qubit addressability in a given field of the governing Hamiltonian, as well as relies on some theoretical simplifications to the true physics of a neutral atom device, highlights a potential solution of this long-known problem. Furthermore, we present how this method along side the refocusing algorithm allows for approximating a wide variety of two-qubit gates in a purely analog fashion. Figure~\ref{fig_intro} shows a schematic of the process outlined in this article. While there are additional steps required in order for the methods we develop to be directly ready for experimental implementation, this work acts as the first step to bridge the gap between analog and digital quantum computing and potentially allows purely digital algorithms to be implemented in an analog quantum device.

This paper is structured as follows: In Section~\ref{sec:rydberg} we introduce the Hamiltonian that governs an analog neutral atom device. In Section~\ref{sec:main} we outline the  main result of this work, i.e.\ a method to approximate layers of single-qubit gates using only global pulses and single qubit addressibility in a single field. In Section~\ref{sec:2qubit_gates} we show, given the access to layers of single-qubit gates, how to perform a refocusing algorithm and thus approximate two-qubit gates. In Sections~\ref{sec:swap} and \ref{sec:vqe} we exemplify the use of these methods through preparing an approximate SWAP gate network and solving simple chemistry problems. In Section~\ref{sec:experiment} we detail further the experimental considerations that need to be taken into account to realise this protocol. Lastly, in Section~\ref{sec:conclusion} we summarise the work presented in this report and outline open questions for future work.

\section{\label{sec:rydberg} Rydberg atoms in tweezers}

In this article we concentrate on the analog capabilities of neutral atom devices. Such systems have attracted much interest for research purposes in many-body physics and more recently for potential applications in industrially relevant use cases. It has been shown that the Maximum Independent Set problem can be solved directly through a mapping to neutral atoms~\cite{pichler2018quantum,wurtz2022industry}; Many machine learning schemes have been proposed that utilise neutral atoms~\cite{PRXQuantum.3.030325,Henry_2021, leclerc2022financial,PhysRevA.107.042615}; A blueprint for VQE protocols taking advantage of the freedom of lattice geometry in registers of neural atoms was recently published~\cite{PhysRevA.107.042602}; Recently, it was also shown how neutral atoms can be used for drug discovery~\cite{darcangelo2023leveraging}. 

In neutral atom devices a qubit is encoded into two electronic states of an atom. Typical choices for atomic species are rubidium or strontium atoms, in which two energy levels can be chosen to represent the two level system of a qubit. Since the number of electronic states for a given atom is large, the various possible choices of $\ket{0}$ and $\ket{1}$ can lead to different interaction landscapes. Throughout this work we will consider encoding $\ket{0}$ in a low-lying energy level and encoding $\ket{1}$ in a Rydberg state; the resultant Hamiltonian of the system with this choice is of Ising type~\cite{henriet2020quantum}
\begin{multline}
    \label{eq_ising}
        \mathcal{H}(t) =\hbar \sum\limits_j \frac{\Omega(t)}{2} \left\{\cos\left[\Phi\left(t\right)\right]\hat{\sigma}_j^x - \sin\left[\Phi\left(t\right)\right]\hat{\sigma}_j^y\right\} 
        \\
        - \hbar  \sum\limits_j \frac{\delta_j (t)}{2} \hat{\sigma}_j^z +\sum\limits_{i > j} \frac{C_6}{r_{ij}^6} \hat{n}_i \hat{n}_j,
\end{multline}
where $\hat{\sigma}^{x,y,z}_i$ are the Pauli $x$, $y$ and $z$ matrices acting on the $i^{\text{th}}$ qubit, $\hat{n}_i=\frac{1-\hat{\sigma}_i^z}{2}$ is the number operator, $r_{ij}$ is the distance between the qubits $i$ and $j$, $\phi$ is the phase, $\Omega$ is the Rabi frequency and $\delta$ the detuning of the external laser that couples the qubit ground and Rydberg states. These can be seen as an effective magnetic field, with transverse and longitudinal components $\propto \Omega(t)$ and $\propto \delta(t)$ respectively. In this work we allow for the detuning to be site dependant. Each field can be varied by changing the intensity and frequency of the laser field. The third term in Eq.~\eqref{eq_ising} relates to the interactions between individual atoms. More specifically, it corresponds to an energy penalty that two qubits experience if they are both in the Rydberg state at the same time - this leads to the well known Rydberg blockade~\cite{Urban_2009}. This coupling between two atoms depends on the coefficient $C_6$ which is defined by the choice of Rydberg state, $n$.

\section{\label{sec:main} Realising analog single-qubit gates}

In this section we propose a method to engineer layers of single-qubit gates on a purely analog neutral atom device, i.e., in the presence of always-on interactions described by Eq.~\eqref{eq_ising}. Given the local addressibility of detunings in Eq.~(\ref{eq_ising}) coupled with the fact that the effect of detuning in the Hamiltonian commutes with the qubit interactions, the implementation of a single qubit Rz gate is quite natural. We first show how these can be routinely realised. 

In order to create pulse sequences (sequences of time-evolution under the Rydberg Hamiltonian with different values of $\Omega$ and $\delta$) that are equivalent to single qubit Rx and Ry gates, we turn to global $\pi/2$ rotations that allow us to change freely between the $X$, $Y$ and $Z$ basis. Given that the strength of the Rabi frequency and detuning are on the same order as the interaction strength between atoms, the protocols that will achieve this are non-trivial. However, in the following we show how a surprisingly simple setup can be used to variationally prepare experimental pulse sequences that approximate these rotations with high accuracy. 

Once these global rotations are obtained, they can be combined with local Rz gates to implement layers of Rx, Ry gates, see Fig.~\ref{fig_pulse_opt}(a). For example,
\begin{equation} \label{eq:local_x}
    \text{Rx}\left(\vec{\theta}\right) \equiv \text{RY}\left(\frac{\pi}{2}\right) \otimes \text{Rz}\left(\vec{\theta}\right) \otimes \text{RY}\left(-\frac{\pi}{2}\right),
\end{equation}
\begin{equation} \label{eq:local_y}
    \text{Ry}\left(\vec{\theta}\right) \equiv \text{RX}\left(\frac{\pi}{2}\right) \otimes \text{Rz}\left(\vec{\theta}\right) \otimes \text{RX}\left(-\frac{\pi}{2}\right).
\end{equation}
where $\vec{\theta}$ is a vector of angles in which each qubit will rotate. Thus, the combination of these rotations and Rz gates allows one to then perform very general single-qubit gates. Note, in the above and throughout this work we will use the notation RX for a global rotation and Rx for a local rotation (and equivalently for $y$ and $z$ rotations).

\subsection{Choice of the geometry}
While the procedure we outline in the following section method can be adapted for more general lattice geometries, as a first example we consider a linear chain of $N$ qubits with nearest neighbour (NN) separation $r$ and periodic boundary conditions (PBC). Furthermore, we neglect next-nearest interactions as they are 64 times smaller than the nearest neighbour interactions. As a result, the interaction term in our Hamiltonian simplifies to
\begin{equation}
J\sum\limits_{i}\hat{n}_i \hat{n}_{i+1},
\end{equation}
where $J=\frac{C_6}{r^6}$. However, in Section.~\ref{sec:experiment} we consider the effect of the full long-range tail of the qubit interactions on the fidelities of the pulse sequences we variationally prepare in the NN setting.

\subsection{Local Rz gates}

A direct consequence of the NN approximation is that we can easily realise any Rz rotation through the single-qubit addressibility in detuning. For $\Omega = 0$, the evolution operator is given by 
\begin{equation}
U(t, \vec{\delta}) \equiv e^{-i\mathcal{H}(\vec{\delta})t},
\end{equation}
where 
\begin{equation}
    \label{eq_rz}
        \mathcal{H}(\vec{\delta}) = J\sum\limits_{i} \hat{n}_i \hat{n}_{i+1} - \hbar \sum\limits_i \frac{\delta_i}{2} \hat{\sigma}_j^z.
\end{equation}
Given that the longitudinal field induced by the laser detuning commutes with the qubit interactions, the effect of these interactions in the Hamiltonian evolution can be removed by carefully choosing the evolution time $t$ such that $tJ = 2\pi$ and thus the effect of $\exp(-itJ\sum_i\hat{n}_i\hat{n}_{i+1}) \equiv \exp(-i2\pi\sum_i\hat{n}_i\hat{n}_{i+1})$ is equivalent to the identity. Thus, we are left with $U(2\pi/J, \vec{\delta}) = \exp\left(i\frac{2\pi}{J}\hbar \sum\limits_i \frac{\delta_i}{2} \hat{\sigma}_i^z\right)$ and, by choosing $\delta_i=\frac{\theta_i}{t}$, this can be tuned to act as Rz gates acting on each qubit with angles $\theta_i$. 

\subsection{Global RX/RY gates}

As shown in Eqs.~\eqref{eq:local_x} and \eqref{eq:local_y}, the combination of local Rz rotations with global $\pi/2$ rotations allows for the realisation of single qubit Rx and Ry rotations. In the following we present the procedure to variationally prepare these global rotation gates, $\left\{\text{RX}\left(\frac{\pi}{2}\right),\text{RX}\left(-\frac{\pi}{2}\right),\text{RY}\left(\frac{\pi}{2}\right),\text{RY}\left(-\frac{\pi}{2}\right)\right\}$, in an analog setting. 

\textbf{Problem setup --} We would like to find a pulse sequence that approximates a global $\text{R}\left(\frac{\pi}{2}\right)$ by optimizing a \textit{global} Rabi frequency, \textit{global} phase and \textit{global} detuning of each pulse, within the constraints $\hbar\Omega<J$ and $\hbar|\delta|<2J$ to ensure experimental feasibility~\cite{wurtz2023aquila, henriet2020quantum}, to minimise a well-chosen loss function. In the following we consider a 4-qubit periodic system with NN interactions. We consider the loss function as the infidelity, $L=1-F$, such that
\begin{equation}\label{eq:fid}
F= \frac{1}{\mathrm{dim}(G)}\left|
\mathrm{tr}\left(G^{\dagger}\prod\limits_{j=1}^{2^p} \exp\left(-i\frac{\Delta t}{\hbar}H(\boldsymbol{\Theta}_j)\right)\right)\right|,
\end{equation}
where $G$ is the desired unitary operation acting on $N$ qubits, in our case $G = \text{R}\left(\frac{\pi}{2}\right)$, $2^p$ is the number of pulses considered, $\Delta t = T/2^p$ in which $T$ is the time length of the full pulse sequence, $\boldsymbol{\Theta}_j=(\Omega_1,\Phi_1,\delta_1, \cdots, \Omega_{2^p},\Phi_{2^p},\delta_{2^p}) $ and $H$ is the Ising Hamiltonian described by Eq.~\eqref{eq_ising} restricted to NN interactions and strictly global pulses. Note that $L = 0$ implies that the pulse sequence found is correct up to a global phase shift.

\textbf{Optimization --} We begin by minimizing the loss function $L$ for a single pulse, i.e. $p=0$, of length $T$ using an L-BFGS-B optimizer. We then use the parameters found with this first optimisation process as an input to a second round of optimisation with two pulses, both of length $T/2$. The effect of this is to increase the expressibility of the pulse sequence. We repeat this, effectively halving each pulse in the sequence split into two individual pulses at each step, until we reach a pulse length $dt$ for each pulse; a predetermined minimum pulse length set by experimental considerations. The full algorithm used is given in Algorithm~\ref{alg_global_rotations}.

Note, given each global rotation can be transformed into any other via a constant shift in the values found for $\phi$, this procedure only needs to be performed once.


\begin{algorithm}
\caption{Global rotation pulse}
\label{alg_global_rotations}
\begin{algorithmic}
    \State $p\mathrm{_{max}}$, $dt \gets$ initialized by user
    \State $T \gets 2^{p\mathrm{_{max}}}dt$
    \State thresh $\gets$ threshold initialized by user
    \State $L$ $\gets$ initialized to a value $>$ thresh
    \While{$L>\text{thresh}$}
        \State $\boldsymbol{\Theta}^{(0)} \gets$ initialized to random values
        \State $t \gets T$
        \For{$p$ in $0:p\mathrm{_{max}}$} 
            \State $L,\boldsymbol{\Theta}^{(p+1)} \gets \min\limits_{\boldsymbol{\Theta}^{(p)}}(L)$
            \State $t \gets T/(2^{p})$
        \EndFor
    \EndWhile
\end{algorithmic}
\end{algorithm}

\begin{figure}
    \centering
     \includegraphics{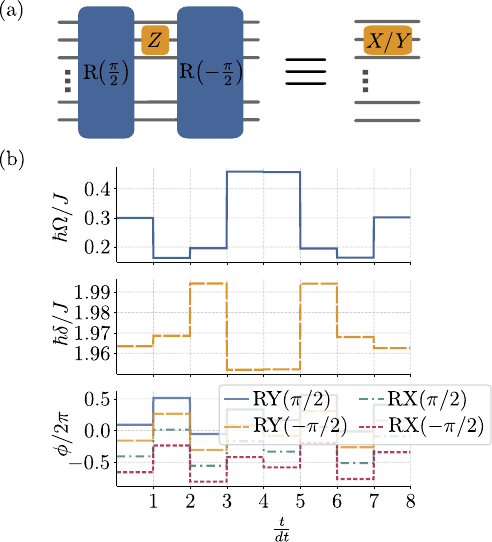}
    \caption{Results of procedure to find a global $\pi/2$ rotation using a neutral atom setup. (a) A schematic showing how these global pulses can be combined with local $Z$ gates to perform local $X$ or $Y$ gates. (b) The precise pulse sequences found to implement global rotations, here $dt\sim 1.6/J$ .}
    \label{fig_pulse_opt}
\end{figure}

\textbf{Results --} We show the results of the procedure outlined above in Fig.~\ref{fig_pulse_opt}(b), here we present the exact pulse sequences found through this procedure that realise all $\pm\pi/2$ rotations with $> 99.9\%$ fidelity for a chain with PBC. The fidelities found for the implementation of these pulses on different qubit systems are presented in Table~\ref{tab_fids_8_lay}.  While this optimisation was performed on a small system, as we consider global pulses and work in the limit of negligible next-nearest neighbour interactions, these pulses generalise well to larger chains. We clearly see that using this measure of fidelity we obtain very accurate results even when doubling the number of qubits in the chain with respect to the number in which the optimisation procedure was implemented on. Counterintuitively, even on such small systems the fidelities of these pulses are even higher when we switch from PBC to open boundary conditions (OBC).

\begin{table}
    \centering
    \begin{tblr}{ colspec= {|c|c|c|c|c|}, hlines = {}, column{even} = {lightgray},row{1}={white}  }
 $\mathrm{n_{qubits}}$ & NN PBC & NN OBC \\
  4 & $99.92\%$ & $99.97\%$ \\
  5 & $99.90\%$ & $99.94\%$ \\
  6 & $99.87\%$ & $99.92\%$ \\
  7 & $99.85\%$ & $99.90\%$ \\
  8 & $99.83\%$ & $99.88\%$ \\
\end{tblr}
    \caption{Fidelities of the global analog qubit rotations for a NN interacting linear chain with PBC and OBC.}
    \label{tab_fids_8_lay}
\end{table}

\begin{figure*}
    \centering
    \includegraphics[width=1.55\columnwidth]{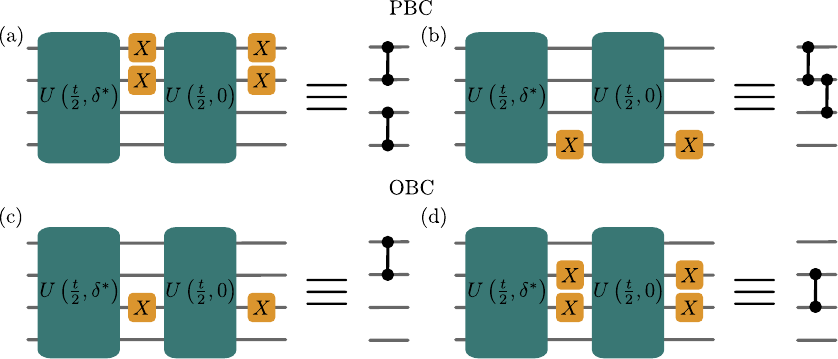}
    \caption{The pulse sequence required to perform a $CZ(0,1)CZ(2,3)$ gate (a) and $CZ(0,1)CZ(1,2)$ gate (b) on a 4 qubit nearest neighbor linear chain with PBC. The pulse sequence required to perform a $CZ(0,1)$ gate (c) and a $CZ(1,2)$ gate (d) on a 4 qubit nearest neighbor linear chain with OBC.}
    \label{fig_engineered_con}
\end{figure*}

\section{Realizing Analog two-qubit gates}\label{sec:2qubit_gates}
Building on the previous section, we now discuss how to realize two-qubit gates with the help of analog rotations. To this end, we first discuss how to `turn off' interactions between qubits with a refocusing algorithm to realise some two-qubit gates~\cite{JONES1999322, PhysRevA.61.042310, PhysRevA.65.040301, Guseynov_2022}. The fundamental idea behind the refocusing algorithm is: by evolving a system under a given pulse, followed by a second pulse of the same length in which the sign of the unwanted interactions is reversed, the resultant effect of this entire pulse sequence is only that of the desired interactions throughout the whole evolution~\cite{Vandersypen05}.

In general reversing the interaction strength of a given quantum system is not physical but this can be effectively achieved for an Ising-type Hamiltonian by using single qubit $X$ gates,
\begin{equation}
    X_ie^{t\sigma^z_i\sigma^z_j}X_i = e^{-t(\sigma^z_i\sigma^z_j)},
\end{equation}
which relies on temporarily flipping the computational basis from $\{0,1\}$ to $\{1,0\}$: in that basis the Pauli-Z operator acquires a sign flip~\cite{parra2020digital}.

Given the close relationship between Rydberg interactions and Ising interactions, and given that layers of $X$ gates can be engineered through the pulse sequences outlined in the previous sections, we are able to realise an analog implementation of the refocusing algorithm. Effectively, this idea will allow us to entangle subsections of the qubit register in a controllable manner. 

\subsection{From NN interaction to $CZ$ gates}
As the most natural example for an analog neutral atom device we will present the procedure to perform different layers of effective $CZ$ gates in an analog operation. A $CZ$ gate acting on two qubits, $i$ and $j$, may be defined as
\begin{equation}
    CZ = \exp\left(i\pi \hat{n}_i\hat{n}_j\right).
\end{equation}
From this definition we can directly recognise the evolution of the nearest-neighbour neutral atom device on an $N$ qubit linear chain with PBC and no external laser fields as
\begin{equation}\label{eq:cz}
    U(t,\vec{0}) = CZ(0,1)CZ(1,2)\dotsb CZ(N,0),
\end{equation} 
providing $t = \pi/J$. Once this equivalence has been established we can consider using the refocusing algorithm to remove the unwanted $CZ$ gates from Eq.~\eqref{eq:cz}. More explicitly, given that the effect of $X$ gates surrounding such a pulse is
\begin{equation}\label{eq:refocus}
\begin{split}
    &X_je^{t\sum_i\hat{n}_i\hat{n}_{i+1}}X_j \\
    &= e^{t\left(\sum_{i\neq j, j-1}\hat{n}_i\hat{n}_{i+1} - \hat{n}_j\hat{n}_{j+1} - \hat{n}_{j-1}\hat{n}_{j} +\hat{n}_{j-1}+\hat{n}_{j+1}\right)}.
\end{split}
\end{equation}
By combining these two results and utilising single qubit detuning to compensate for the linear term in the exponent of Eq.~\eqref{eq:refocus}, we are able to produce layers of $CZ$ gates between desired qubits. Note, to `turn off' the interaction between qubits $i$ and $j$, we surround the second pulse in the sequence by $X$ gates acting on one of the two qubits involved in the interaction we wish to turn off. When no $X$ gates act on the qubit pair or an $X$ gate acts on both qubits, the interaction will not be affected.  For example, to remove the $CZ(0,1)$ and $CZ(2,3)$ from Eq.\eqref{eq:cz} we can perform the pulse sequence
\begin{equation}
\begin{split}
    &CZ(1,2)CZ(3,4)CZ(4,5)\dotsb CZ(N,0) = \\
    &X_1X_2 \otimes U\left(\frac{t}{2}, \vec{0}\right) \otimes X_1X_2 \otimes U\left(\frac{t}{2}, \vec{\delta}^*\right),
\end{split}
\end{equation}
where $\vec{\delta}^*$ is chosen to account for the unwanted linear terms accumulated by the action of each $X$ gate.

The types of layers of CZ gates realisable through this protocol of removing given interactions depends on the geometry and boundary conditions of the system. In the following we will explain these explicitly for a linear chain with both PBC and OBC.

\subsection{Periodic Boundary Conditions: Even and odd qubit chains} 
Given that the action of each $X$ gate is to remove the effect of two distinct interactions in a system with PBC, it becomes clear that the number of qubits in the chain, even or odd, defines the subset of $CZ$ gates realisable. Explicitly, in an even-numbered qubit chain we are able to realise layers of an even number of $CZ$ gates, and in an odd-numbered qubit chain we are able to realise layers of an odd number of $CZ$ gates. As a result, in order to realise a single $CZ$ gate, for example $CZ(0,1)$, we are required to work on a system containing an odd number of qubits. However, as many applications benefit from parallelisation of two qubit gates, this does not greatly restrict potential use cases of our method.

As an explicit example we consider a 4 qubit chain with PBC. The possible two types of layers of $CZ$ gates realisable (up to transnational symmetry) are $CZ(0,1)CZ(2,3)$, by, for example, using $X$ gates on qubits 0 and 1 and $CZ(0,1)CZ(1,2)$ by, for example, using an $X$ gate on qubit 3. Schematics of these pulse sequences are given in Figs.~\ref{fig_engineered_con}(a) and (b), each requiring pulse sequences of length $4T+5\pi/J$. These achieve a fidelity of $99.6\%$ and $99.7\%$ respectively for a system with NN interaction. 

\subsection{Open Boundary Conditions}

Given a linear chain with OBC we can readily prepare a given layer of CZ gates with no restrictions - this is a consequence of the end qubits interacting with only one other qubit. For example, we an prepare a $CZ(0,1)$ gate on through the use of an $X$ gate on qubit 2 or a $CZ(1,2)$ through the use of $X$ gates on qubits 1 and 2. Schematics of these pulse sequences are given in Figs.~\ref{fig_engineered_con}(c) and (d), again each require pulse sequences of length $4T+5\pi/J$. These achieve a fidelity of $99.7\%$ and $99.6\%$ respectively for a system with NN interaction. 

Given that the variationally generated pulse sequences for global rotations have higher fidelity on linear chains with OBC as well as the additional flexibilty to realise layers of CZ gates this setting gives, we will consider such systems for the rest of this article. In the following sections, we show how, in theory, analog single qubit and $CZ$ gates can be used to build a full SWAP network to produce all-to-all connectivity in an analog neutral atom device. We then use this result to perform some simple analog VQE simulations of small molecules.

\section{\label{sec:swap} Generating a SWAP network}

While some attempts have been made to build devices with all-to-all connectivity (for instance using trapped ions~\cite{PhysRevA.100.022332}, photons~\cite{bartolucci2021fusionbased} as well as some preliminary results in neutral atoms~\cite{Bluvstein_2022,Bluvstein_2023}) in currently available NISQ devices based on superconducting qubits~\cite{patra2023efficient,pan2021simulating} and neutral atoms~\cite{henriet2020quantum} connectivity is limited. An example in which this low connectivity can be an issue is the simulation of fermionic systems. When simulating the physics of such systems, the occupation number of a spin orbital is encoded in the qubit state. Therefore, in order to allow all orbitals to interact with each other we must apply two qubit gates between all pairs of qubits. Mathematically, this can be seen through the loss of locality when applying the Jordan-Wigner transformation on simple molecular Hamiltonians to convert between the fermionic basis and qubit basis. Using quantum annealing to solve certain difficult quadratic unconstrained binary optimization (QUBO) problems is another example in which all-to-all connectivity is required~\cite{D_ez_Valle_2023}. Such connectivity allows many NP-hard optimisation problems to be encoded in the ground state of an Ising model. This is thought to be a promising direction for useful quantum computation~\cite{Lucas_2014}.

\begin{figure}
    \centering
    \includegraphics{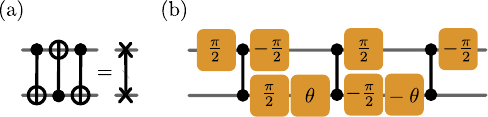}
    \caption{(a) The decomposition of a SWAP gate as three CNOT gates. (b) The decomposition of a Givens SWAP gate into $CZ$ gates and single-qubit Ry rotations, note the angles of each rotation are shown in the schematic.}
    \label{fig:gate_decomp}
\end{figure}

One popular method to circumvent the issue of low connectivity within a digital device is a SWAP network. Such a network uses SWAP gates that switch the quantum information between two given qubits. A SWAP network effectively inverts the position of the qubits in the circuit and allows one to act the required operations between the given qubits when they are adjacent to each other in this cycle~\cite{ogorman2019generalized}. Clearly, performing such a SWAP network is not native to an analog device. Typically, in analog systems long-range entanglement is generated through quench dynamics and is limited by a Lieb-Robinson bound of the resource Hamiltonian~\cite{PhysRevA.107.042602}. However, using the protocols outlined in the Sections~\ref{sec:main} and \ref{sec:2qubit_gates}, a SWAP network can be directly imported into the analog setting. In particular, a SWAP gate can be decomposed into three CNOT gates as shown in Fig.~\ref{fig:gate_decomp}(a). As a CNOT gate can be realised through
\begin{equation}
    CNOT(i,j) = \text{Ry}\left(\frac{\pi}{2}\right)\;CZ(i,j)\;\text{Ry}\left(-\frac{\pi}{2}\right),
\end{equation}
we conclude that with the pulse sequences given in this work, an analog SWAP gate is, in principle, realisable.

As a simple benchmark, by using the gates found in the previous sections to perform a SWAP network on 4 qubits in a NN linear chain, we realise a long-range CNOT gate between qubits 0 and 3. This pulse sequence achieved a fidelity of $~94\%$ with SWAP gates that each require pulse sequences of length $3(6T+5\pi/J)$. In the following, we add one more layer of complexity to this proposal by replacing SWAP gates with Givens SWAP gates that are known to be an effective ansatz for VQE problems~\cite{Elfving2021}. 

\section{\label{sec:vqe} Application: Variational Quantum Eigensolvers} 
As previously mentioned, one promising application of quantum computing that strongly benefits from all-to-all connectivity is the simulation of molecules. When simulating the physics of electronic systems the typical approach is to use Fock basis in which each basis function represents a fermionic mode. A standard method to simplify the task of simulating these systems as well as half the required resources is the paired-electron approximation. Here, we consider the simplification such that each molecular orbital is either occupied by an electron singlet pair or unoccupied. Under this assumption the Hamiltonian of a molecule is given by 
\begin{equation}
    \hat{\cal{H}}=C + \sum_{p,q}h_{p,q}^{(r1)}\hat{b}_p^{\dagger}\hat{b}_q+\sum_{p \neq q}h_{p,q}^{(r2)}\hat{b}_p^{\dagger}\hat{b}_p\hat{b}_q^{\dagger}\hat{b}_q,
\end{equation}
where $\hat{b}_q$ are hardcore bosons representing the electron-pair annihilation operator in mode $p$ that satisfy the relevant (anti-)commutation relations and $h_{p,q}^{(r1)}$ and $h_{p,q}^{(r2)}$ can be calculated from the one- and two-electron integrals~\cite{Elfving2021}.

To map this problem to a form that is applicable for a quantum computer we can use the transformation $\hat{b}_p=\frac{1}{2}(\hat{\sigma}^x_p+i\hat{\sigma}^y_p),$
resulting in a Hamiltonian of the form
\begin{multline}
\label{eq:pair_electron}
   \hat{\cal{H}}_{qb}\approx C + \sum_p \frac{h_p^{(r1)}}{2}(\hat{I}_p-\hat{\sigma}_p^z) +\sum_{p \neq q} \frac{h_{p,q}^{(r1)}}{4}(\hat{\sigma}_p^x\hat{\sigma}_q^x+\hat{\sigma}_p^y\hat{\sigma}_q^y)\\+\sum_{p \neq q}\frac{h_{p,q}^{(r2)}}{4}(\hat{I}_p-\hat{\sigma}_p^z-\hat{\sigma}_q^z+\hat{\sigma}_p^z\hat{\sigma}_q^z).
\end{multline}
This form of the Hamiltonian is now in a basis that directly maps to qubits and allows us to find the ground state energy through a Variational Quantum Eigensolver (VQE) approach.

\begin{figure}
    \centering
    \includegraphics{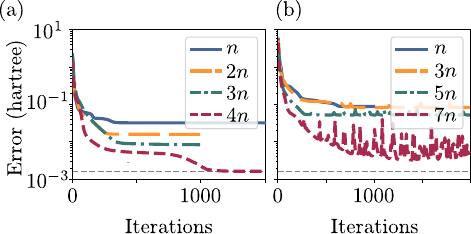}
    \caption{Analog VQE results for the molecules H$_2$ in (a) and LiH in (b), both run in Qadence~\cite{seitz2024qadence}. Here, we show the results for multiple ansatzes with varying depths of Given SWAP gates as a function of the number of qubits, $n$. For H$_2$ $n=4$ and LiH $n=6$. Here we manage to reduce the error of the VQE protocol to be on the same order of magnitude of the chemical accuracy (the grey horizontal line).}
    \label{fig:VQE_results}
\end{figure}

Given the molecular Hamiltonian is number conserving, i.e., $\sum_{p=0}^q \hat{b}_p^{\dagger}\hat{b}_q=n$ where $n$ is the number of electrons in the system, a good choice for initial state during a VQE protocol is the Hartree-Fock state. This is the state in which the $n$ lowest energy molecular orbitals are occupied with pairs of electrons, here $n$ is the total number of pairs of electrons in the system. From this initial state a Givens SWAP network has been shown to be a promising variational ansatz to find the ground state within the paired-electron approximation~\cite{Elfving2021}. In fact, a proof of principle experiment utilising this ansatz was preformed on Google's superconducting qubit quantum computer and successfully found the ground state for some simple molecular Hamiltonians~\cite{obrien2022purificationbased}. A Givens SWAP gate is the combination of a Givens rotation followed by a SWAP gate, mathematically it is defined by
\begin{equation}
    GS(\theta)= 
    \underbrace{
    \begin{pmatrix}
        1 & 0 & 0 & 0 \\
        0 & 0 & 1 & 0 \\
        0 & 1 & 0 & 0 \\
        0 & 0 & 0 & 1
    \end{pmatrix}
    }_{SWAP}
    \underbrace{
        \begin{pmatrix}
            1 & 0 & 0 & 0 \\
            0 & \cos(\theta) & -\sin(\theta) & 0 \\
            0 & \sin(\theta) & \cos(\theta) & 0 \\
            0 & 0 & 0 & 1
        \end{pmatrix}
    }_{G(\theta)}
\end{equation}
in which $\theta$ is a variational parameter. A full digital Givens SWAP network is equivalent to the SWAP network with each SWAP gate replaced by a Givens SWAP gate.

We now turn our attention to the analog implementation of such an ansatz. As a Givens SWAP network can be decomposed into three $CZ$ gates and layers of single-qubit gates, see Fig.~\ref{fig:gate_decomp}(b), we can again lean on the protocols outlined in Sections~\ref{sec:main} and \ref{sec:2qubit_gates} to build an analog pulse sequence that approximates this digital circuit.

We use such a variational architecture to find the ground state energy of H$_2$ and LiH using an analog Givens SWAP network. For H$_2$ we consider the 631g basis in which 4 qubits are required in the paired-election approximation~\cite{jensen2017introduction}. For LiH we consider the sto-3g basis in which we require 6 qubits in the paired-electron approximation~\cite{jensen2017introduction}. In Fig.~\ref{fig:VQE_results} we show the evolution of errors throughout the VQE protocol. We define the error as the difference between the measured energy and the lowest value achievable in the paired electron approach. Theoretically, the number of layers of Givens SWAPs required to reach the ground state is equal to the number of qubits. However, since our gates do not have a $100\%$ fidelity, more layers are required. We vary the depth of our network as a function of the number of paired electrons $n$, i.e. $n=4$ for H$_2$ and $n=6$ for LiH. Here, we used BFGS, a gradient based optimizer and an a convergence criterion of $\epsilon = e^{-7}$. In Fig.~\ref{fig:VQE_results} see that while chemical accuracy (an error of 1.6 mHa) was not achieved for either H$_2$ of LiH, we were to achieve an error on the same order of magnitude to this goal.

\section{\label{sec:experiment}Experimental Consideration}
In this section we consider realistic experimental constraints such as atomic positions, interactions strengths, Rabi frequencies, detunings and coherence times. From these we present the required coherence times to realise each gate used throughout the results we present in this report. Furthermore, we show the effect of the long-range nature of interactions in a realistic qubit register on the accuracy of the pulses.

\subsection{Pulse sequence depths}
We consider a rubidium-based neutral atom device with atoms coupled to the $|60S_{1/2}\rangle$ Rydberg state, and include the full $1/r^6$ long-range tail of the Van der Waals interactions. Furthermore, we consider $N$ atoms in a circular geometry with inter-atomic spacing of $r=6.24~\mu$m. This results in a nearest neighbour interaction of $J \sim 2\pi \times 2.3$ MHz. We consider $dt = 108$ ns such that the time to perform a local $RZ$ gate is $428$ ns with $\left|\delta_j\right| = J$. Our variationally prepared pulse sequences for the global $\pi/2$ rotation gates are such that $T=864$ ns, $\Omega/2\pi<1.2$ MHz and $\left|\delta/2\pi\right|<4.6$ MHz. From these values we can then calculate the required pulse lengths of each of the gates we use in this work, see Table.~\ref{tab_pulse_lengths}. While the analog SWAP and Givens SWAP gates may be beyond current experimental limitations with the pulses presented here, given that coherence times of up to $\sim\!6\mu$s have been experimentally realised~\cite{Scholl_2021}, it is not unreasonable to expect a layer of analog CNOT gates to be feasible in the near future since this coherence time can be improved by further stabilizing some of the control components on the hardware. Alternatively, the involved energy scales could be increased uniformly, speeding up all of the operations described here. Ultimately the limiting timescale is the Rydberg lifetime, which is on the order of 100 $\mu$s for the state chosen here. However, this too could be extended by choosing a larger principal quantum number. Note, in order to run deep quantum circuits with more than one unique layer of qubit gates, we would require temporal control of the single qubit addressability in laser detunings. 

\begin{table}
    \centering
    \begin{tblr}{ colspec= {|c|c|}, hlines = {}, column{even} = {lightgray},row{1}={white}  }
(layer of) Gate & Pulse Sequence Length ($\mu$s) \\
  RZ & $0.4$ \\
  Global $\pi/2$ rotation & $0.9$ \\
  Local RX/RY &  $2.1$ \\
  CZ & $4.5$  \\
  CNOT & $6.2$ \\
  SWAP & $18.7$ \\
  Givens SWAP & $22.1$
\end{tblr}
    \caption{The required pulse lengths of given gates using proposed analog pulses.}
    \label{tab_pulse_lengths}
\end{table}

\subsection{Effect of long-range interactions}\label{sec:long-range}
In the following section we consider the effect of the full long-range interaction tail on the fidelities of both our proposal for local Rz gates proposed in Section.~\ref{sec:main} as well as the pulses given in Fig.~\ref{fig_pulse_opt}(b). To estimate the effect of these long-range interactions in the limit $N\gg1$ we consider a linear spin chain with OBC.

While the protocol to produce a local Rz gate proposed in Section.~\ref{sec:main} is exact for NN interactions, it is only approximate in the presence of the full long-range interaction tail defined in Eq.~\eqref{eq_ising}. As a result we find that the fidelity drops to $~98.9\%$ for a system of 8 qubits. However, by deviating the pulse length used away from the theoretically exact time for a NN system we are able to retain a fidelity of $99.6\%$, see Fig.~\ref{fig:Exp_results}(a). Furthermore, in this setting of 8 qubits, the fidelity of the global rotation pulse sequences falls by about $2-3\%$. However, by again allowing the length of the pulse sequence to vary we can obtain a fidelity of $98.7\%$, see Fig.~\ref{fig:Exp_results}(b).

While these fidelities are still high and may be acceptable for certain applications, it highlights that, given access to more computational resources, performing the optimisation protocol outlined above on a larger system with the full long-range tail would generalise better to long chains.

\subsection{Generalisation to larger system sizes}
\begin{figure}
    \centering
    \includegraphics{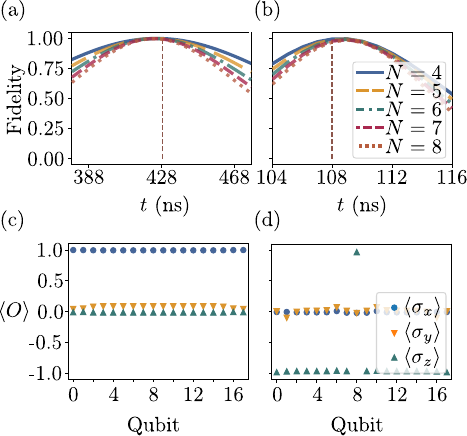}
    \caption{Full analog simulations of both the global rotations as well as a single qubit $X$ gate using the full long-range tail of the Van der Waals interactions. (a) and (b) Results for the fidelity of analog Rz and global rotation gates respectively in the presence of long-range interactions showing the effect of varying the pulse length from the analytically predicted value using NN interactions (shown as a vertical dashed line). By varying the pulse lengths we are able to boast the fidelities of the analog Rz and global rotation gates to $99.6\%$ and $98.7\%$ respectively. (c) and (d) The resultant local magnetisations after the analog pulses for a global $\text{RY}\left(\frac{\pi}{2}\right)$ and local $X_8$ gates respectively.}
    \label{fig:Exp_results}
\end{figure}

In Fig.~\ref{fig:Exp_results} we show how our protocol generalises to large qubit systems, $N=18$ which is more than four times the chain length the pulses were optimised over. In these simulations we consider the full long-range tails of the Van der Waals interactions, here we place the qubits in a linear chain with OBC. As the fidelity given in Eq.~\eqref{eq:fid} requires the calculation of the full unitary time evolution operator for a given pulse sequence, it is an expensive metric for chains of this length.  Thus, we use a simplified metric for fidelity: the overlap the state prepared by acting our proposed pulse sequence on the ground state with the exact state that is prepared via the target gate, $G$, i.e.
\begin{equation}
    \Tilde{F} = \bra{000\cdots0} G^{\dagger}\prod\limits_{i=1}^p \exp\left(-i\frac{\Delta t}{p \hbar}H(\boldsymbol{\Theta})\right)\ket{000\cdots0}.
\end{equation}
We find that for the pulse sequences variationally found for global $\pi/2$ rotations retain an overlap of $\Tilde{F}>97\%$. 

Furthermore, we ran simulations for a local $X$ gate on a single qubit in this experimentally inspired set up with a resulting overlap of $\Tilde{F}\approx85\%$. However, as mentioned previously, when introducing long-range interactions the Rz gate implementation is no longer exact. Thus, as further investigation we find that deviating the pulse length from the NN predicted time to $412$ns, we are able to increase the fidelity of the single qubit Rx gate to $\Tilde{F}>93\%$. In Fig.~\ref{fig:Exp_results}(c) and (d) we show the resultant local magnetisations of each qubit after the analog pulses for both a global $\text{RY}\left(\frac{\pi}{2}\right)$ rotation and local $X$ gate in the bulk of the chain. Both of these results show that our pulse sequences optimised on only 4 qubits with NN interactions are still highly effective in large systems with the full long-range tail, and clearly show (up to small errors) the desired gate effect. 

\subsection{Future directions}
In order to both reduce the required coherence times to implement these analog gates as well as increase the fidelities in a more realistic experimental set up, there are many directions that can be considered. Firstly, by performing the optimisation with more flexible pulse sequences, for example continuous interpolated pulse sequences parametrized through a Fourier basis~\cite{Geen91}, or \textit{bang-bang} pulses~\cite{PhysRevX.7.021027} and by making use of the CRAB algorithm~\cite{Caneva11}), we can aim to reduce $T$. Furthermore, performing the optimisation over a larger number of qubits in a circular register and taking into account the full long-range nature of interactions would push up the fidelities observed in the experimental setup used in this section. 

In order to solve the control task more efficiently, this work could be extended using several other optimization techniques. Classical simulation algorithms that perform gradient-based optimization such as GRAPE~\cite{Khaneja05}, for instance, would offer a drop-in replacement for the approach taken here. In particular, GRAPE also discretizes the control pulses, but then optimizes based on the (approximate) derivative of the cost function with respect to the pulse train parameters. As a purely classical algorithm, it is restricted to systems consisting of a small number of qubits. For larger systems, hybrid approaches such as differentiable analog quantum computing~\cite{Leng22} offer an attractive alternative. Here, the (functional) derivative of the cost function with respect to some control function is calculated on a quantum system using a combination of parameter shift rules and Monte-Carlo integration.

Alternatively, gradient-free techniques such as Bayesian optimization~\cite{Shahriari15, Frazier18} could be helpful. The downside of such an approach, however, is its unfriendly scaling with the dimension of the parameter space it has to search through. This is especially a problem with fine-grained pulses that have a lot of degrees of freedom. Restricting oneself to pulses generated by a small set of basis functions would make such methods attractive.

Experimentally, we did not consider varying both the Rydberg level in use and the inter-atomic spacing. This extra freedom could be utilised to decrease $dt$ and reduce the required pulse length of a given quantum circuit. More generally, experimental improvements that allow increased coherence times, larger laser detuning and Rabi frequencies as well as an increased maximal gradient of laser intensity as a function of time, all would give the protocols developed in this article more flexibility and thus increase their utility.

\section{\label{sec:conclusion} Outlook} 
In this work we outlined a procedure to realise a general set of one- and two-qubit gates in the presence of \textit{always-on} interactions. In particular, we considered simulations using an analog neutral atom device. In this setting, we showed how local Rz rotations are realisable given local addressing on the laser detunings. Furthermore, we presented a procedure to variationally prepare global $\pi/2$ rotations to allow us to change basis between the $X$, $Y$ and $Z$ axis with high fidelity. Combining these methods we are able to perform layers of single-qubit gates. These in turn allowed us to build up a repertoire of two-qubits gates taking advantage of a refocusing algorithm.

While currently using the methods in this work to import a fully digital algorithm into an analog setting is still far beyond what is experimentally feasible, there are many short term uses that are highly valuable. For example, one hurdle for analog simulation on a neutral atom device is initial state preparation. With the analog pulses outlined above, we are able to prepare both $X$ and $Y$ polarised states as well as product states with high fidelities. By generalising this protocol to higher dimensions, while non-trivial, would open up the prospect of more interesting quantum simulations in the near-term. Furthermore, this work opens the door to measurements in a general basis. This would allow for measurements of new observables, for example energy and general correlation functions. There has been some exciting proposals on the use of randomised measurements in the neutral atom setting~\cite{notarnicola2023randomized}. A successful implementation of the above protocols outlined in this work would open up the door for such methods to be implemented in the near term.

\acknowledgements
We thank P. Barkoutsos and D. Seitz for insightful discussions. AD and JV acknowledge funding from the European Union under Grant Agreement 101080142 and the project EQUALITY. JdH acknowledges support from the Horizon Europe programme HORIZON-CL4-2021-DIGITAL-EMERGING-01-30 via the project 101070144 (EuRyQa).

\bibliography{bib}

\begin{thebibliography}{81}%
\makeatletter
\providecommand \@ifxundefined [1]{%
 \@ifx{#1\undefined}
}%
\providecommand \@ifnum [1]{%
 \ifnum #1\expandafter \@firstoftwo
 \else \expandafter \@secondoftwo
 \fi
}%
\providecommand \@ifx [1]{%
 \ifx #1\expandafter \@firstoftwo
 \else \expandafter \@secondoftwo
 \fi
}%
\providecommand \natexlab [1]{#1}%
\providecommand \enquote  [1]{``#1''}%
\providecommand \bibnamefont  [1]{#1}%
\providecommand \bibfnamefont [1]{#1}%
\providecommand \citenamefont [1]{#1}%
\providecommand \href@noop [0]{\@secondoftwo}%
\providecommand \href [0]{\begingroup \@sanitize@url \@href}%
\providecommand \@href[1]{\@@startlink{#1}\@@href}%
\providecommand \@@href[1]{\endgroup#1\@@endlink}%
\providecommand \@sanitize@url [0]{\catcode `\\12\catcode `\$12\catcode
  `\&12\catcode `\#12\catcode `\^12\catcode `\_12\catcode `\%12\relax}%
\providecommand \@@startlink[1]{}%
\providecommand \@@endlink[0]{}%
\providecommand \url  [0]{\begingroup\@sanitize@url \@url }%
\providecommand \@url [1]{\endgroup\@href {#1}{\urlprefix }}%
\providecommand \urlprefix  [0]{URL }%
\providecommand \Eprint [0]{\href }%
\providecommand \doibase [0]{http://dx.doi.org/}%
\providecommand \selectlanguage [0]{\@gobble}%
\providecommand \bibinfo  [0]{\@secondoftwo}%
\providecommand \bibfield  [0]{\@secondoftwo}%
\providecommand \translation [1]{[#1]}%
\providecommand \BibitemOpen [0]{}%
\providecommand \bibitemStop [0]{}%
\providecommand \bibitemNoStop [0]{.\EOS\space}%
\providecommand \EOS [0]{\spacefactor3000\relax}%
\providecommand \BibitemShut  [1]{\csname bibitem#1\endcsname}%
\let\auto@bib@innerbib\@empty
\bibitem [{\citenamefont {Lloyd}(1996)}]{doi:10.1126/science.273.5278.1073}%
  \BibitemOpen
  \bibfield  {author} {\bibinfo {author} {\bibfnamefont {S.}~\bibnamefont
  {Lloyd}},\ }\href {\doibase 10.1126/science.273.5278.1073} {\bibfield
  {journal} {\bibinfo  {journal} {Science}\ }\textbf {\bibinfo {volume}
  {273}},\ \bibinfo {pages} {1073} (\bibinfo {year} {1996})},\ \Eprint
  {http://arxiv.org/abs/https://doi.org/10.1126/science.273.5278.1073}
  {https://doi.org/10.1126/science.273.5278.1073} \BibitemShut {NoStop}%
\bibitem [{\citenamefont {Shor}(1994)}]{365700}%
  \BibitemOpen
  \bibfield  {author} {\bibinfo {author} {\bibfnamefont {P.}~\bibnamefont
  {Shor}},\ }in\ \href {\doibase 10.1109/SFCS.1994.365700} {\emph {\bibinfo
  {booktitle} {Proceedings 35th Annual Symposium on Foundations of Computer
  Science}}}\ (\bibinfo {year} {1994})\ pp.\ \bibinfo {pages}
  {124--134}\BibitemShut {NoStop}%
\bibitem [{\citenamefont {Koch}\ \emph {et~al.}(2007)\citenamefont {Koch},
  \citenamefont {Yu}, \citenamefont {Gambetta}, \citenamefont {Houck},
  \citenamefont {Schuster}, \citenamefont {Majer}, \citenamefont {Blais},
  \citenamefont {Devoret}, \citenamefont {Girvin},\ and\ \citenamefont
  {Schoelkopf}}]{PhysRevA.76.042319}%
  \BibitemOpen
  \bibfield  {author} {\bibinfo {author} {\bibfnamefont {J.}~\bibnamefont
  {Koch}}, \bibinfo {author} {\bibfnamefont {T.~M.}\ \bibnamefont {Yu}},
  \bibinfo {author} {\bibfnamefont {J.}~\bibnamefont {Gambetta}}, \bibinfo
  {author} {\bibfnamefont {A.~A.}\ \bibnamefont {Houck}}, \bibinfo {author}
  {\bibfnamefont {D.~I.}\ \bibnamefont {Schuster}}, \bibinfo {author}
  {\bibfnamefont {J.}~\bibnamefont {Majer}}, \bibinfo {author} {\bibfnamefont
  {A.}~\bibnamefont {Blais}}, \bibinfo {author} {\bibfnamefont {M.~H.}\
  \bibnamefont {Devoret}}, \bibinfo {author} {\bibfnamefont {S.~M.}\
  \bibnamefont {Girvin}}, \ and\ \bibinfo {author} {\bibfnamefont {R.~J.}\
  \bibnamefont {Schoelkopf}},\ }\href {\doibase 10.1103/PhysRevA.76.042319}
  {\bibfield  {journal} {\bibinfo  {journal} {Phys. Rev. A}\ }\textbf {\bibinfo
  {volume} {76}},\ \bibinfo {pages} {042319} (\bibinfo {year}
  {2007})}\BibitemShut {NoStop}%
\bibitem [{\citenamefont {Kielpinski}\ \emph {et~al.}(2002)\citenamefont
  {Kielpinski}, \citenamefont {Monroe},\ and\ \citenamefont
  {Wineland}}]{kielpinski2002architecture}%
  \BibitemOpen
  \bibfield  {author} {\bibinfo {author} {\bibfnamefont {D.}~\bibnamefont
  {Kielpinski}}, \bibinfo {author} {\bibfnamefont {C.}~\bibnamefont {Monroe}},
  \ and\ \bibinfo {author} {\bibfnamefont {D.~J.}\ \bibnamefont {Wineland}},\
  }\href {https://www.nature.com/articles/nature00784} {\bibfield  {journal}
  {\bibinfo  {journal} {Nature}\ }\textbf {\bibinfo {volume} {417}},\ \bibinfo
  {pages} {709} (\bibinfo {year} {2002})}\BibitemShut {NoStop}%
\bibitem [{\citenamefont {Bruzewicz}\ \emph {et~al.}(2019)\citenamefont
  {Bruzewicz}, \citenamefont {Chiaverini}, \citenamefont {McConnell},\ and\
  \citenamefont {Sage}}]{bruzewicz2019trapped}%
  \BibitemOpen
  \bibfield  {author} {\bibinfo {author} {\bibfnamefont {C.~D.}\ \bibnamefont
  {Bruzewicz}}, \bibinfo {author} {\bibfnamefont {J.}~\bibnamefont
  {Chiaverini}}, \bibinfo {author} {\bibfnamefont {R.}~\bibnamefont
  {McConnell}}, \ and\ \bibinfo {author} {\bibfnamefont {J.~M.}\ \bibnamefont
  {Sage}},\ }\href
  {https://pubs.aip.org/aip/apr/article-abstract/6/2/021314/570103/Trapped-ion-quantum-computing-Progress-and}
  {\bibfield  {journal} {\bibinfo  {journal} {Applied Physics Reviews}\
  }\textbf {\bibinfo {volume} {6}} (\bibinfo {year} {2019})}\BibitemShut
  {NoStop}%
\bibitem [{\citenamefont {Henriet}\ \emph {et~al.}(2020)\citenamefont
  {Henriet}, \citenamefont {Beguin}, \citenamefont {Signoles}, \citenamefont
  {Lahaye}, \citenamefont {Browaeys}, \citenamefont {Reymond},\ and\
  \citenamefont {Jurczak}}]{henriet2020quantum}%
  \BibitemOpen
  \bibfield  {author} {\bibinfo {author} {\bibfnamefont {L.}~\bibnamefont
  {Henriet}}, \bibinfo {author} {\bibfnamefont {L.}~\bibnamefont {Beguin}},
  \bibinfo {author} {\bibfnamefont {A.}~\bibnamefont {Signoles}}, \bibinfo
  {author} {\bibfnamefont {T.}~\bibnamefont {Lahaye}}, \bibinfo {author}
  {\bibfnamefont {A.}~\bibnamefont {Browaeys}}, \bibinfo {author}
  {\bibfnamefont {G.-O.}\ \bibnamefont {Reymond}}, \ and\ \bibinfo {author}
  {\bibfnamefont {C.}~\bibnamefont {Jurczak}},\ }\href
  {https://quantum-journal.org/papers/q-2020-09-21-327/} {\bibfield  {journal}
  {\bibinfo  {journal} {Quantum}\ }\textbf {\bibinfo {volume} {4}},\ \bibinfo
  {pages} {327} (\bibinfo {year} {2020})}\BibitemShut {NoStop}%
\bibitem [{\citenamefont {Bluvstein}\ \emph {et~al.}(2023)\citenamefont
  {Bluvstein}, \citenamefont {Evered}, \citenamefont {Geim}, \citenamefont
  {Li}, \citenamefont {Zhou}, \citenamefont {Manovitz}, \citenamefont {Ebadi},
  \citenamefont {Cain}, \citenamefont {Kalinowski}, \citenamefont {Hangleiter},
  \citenamefont {Bonilla~Ataides}, \citenamefont {Maskara}, \citenamefont
  {Cong}, \citenamefont {Gao}, \citenamefont {Sales~Rodriguez} \emph
  {et~al.}}]{Bluvstein_2023}%
  \BibitemOpen
  \bibfield  {author} {\bibinfo {author} {\bibfnamefont {D.}~\bibnamefont
  {Bluvstein}}, \bibinfo {author} {\bibfnamefont {S.~J.}\ \bibnamefont
  {Evered}}, \bibinfo {author} {\bibfnamefont {A.~A.}\ \bibnamefont {Geim}},
  \bibinfo {author} {\bibfnamefont {S.~H.}\ \bibnamefont {Li}}, \bibinfo
  {author} {\bibfnamefont {H.}~\bibnamefont {Zhou}}, \bibinfo {author}
  {\bibfnamefont {T.}~\bibnamefont {Manovitz}}, \bibinfo {author}
  {\bibfnamefont {S.}~\bibnamefont {Ebadi}}, \bibinfo {author} {\bibfnamefont
  {M.}~\bibnamefont {Cain}}, \bibinfo {author} {\bibfnamefont {M.}~\bibnamefont
  {Kalinowski}}, \bibinfo {author} {\bibfnamefont {D.}~\bibnamefont
  {Hangleiter}}, \bibinfo {author} {\bibfnamefont {J.~P.}\ \bibnamefont
  {Bonilla~Ataides}}, \bibinfo {author} {\bibfnamefont {N.}~\bibnamefont
  {Maskara}}, \bibinfo {author} {\bibfnamefont {I.}~\bibnamefont {Cong}},
  \bibinfo {author} {\bibfnamefont {X.}~\bibnamefont {Gao}}, \bibinfo {author}
  {\bibfnamefont {P.}~\bibnamefont {Sales~Rodriguez}},  \emph {et~al.},\ }\href
  {\doibase 10.1038/s41586-023-06927-3} {\bibfield  {journal} {\bibinfo
  {journal} {Nature}\ } (\bibinfo {year} {2023}),\
  10.1038/s41586-023-06927-3}\BibitemShut {NoStop}%
\bibitem [{\citenamefont {Slussarenko}\ and\ \citenamefont
  {Pryde}(2019)}]{slussarenko2019photonic}%
  \BibitemOpen
  \bibfield  {author} {\bibinfo {author} {\bibfnamefont {S.}~\bibnamefont
  {Slussarenko}}\ and\ \bibinfo {author} {\bibfnamefont {G.~J.}\ \bibnamefont
  {Pryde}},\ }\href {https://doi.org/10.1063/1.5115814} {\bibfield  {journal}
  {\bibinfo  {journal} {Applied Physics Reviews}\ }\textbf {\bibinfo {volume}
  {6}} (\bibinfo {year} {2019})}\BibitemShut {NoStop}%
\bibitem [{\citenamefont {Ryan-Anderson}\ \emph {et~al.}(2021)\citenamefont
  {Ryan-Anderson}, \citenamefont {Bohnet}, \citenamefont {Lee}, \citenamefont
  {Gresh}, \citenamefont {Hankin}, \citenamefont {Gaebler}, \citenamefont
  {Francois}, \citenamefont {Chernoguzov}, \citenamefont {Lucchetti},
  \citenamefont {Brown}, \citenamefont {Gatterman}, \citenamefont {Halit},
  \citenamefont {Gilmore}, \citenamefont {Gerber}, \citenamefont {Neyenhuis},
  \citenamefont {Hayes} \emph {et~al.}}]{PhysRevX.11.041058}%
  \BibitemOpen
  \bibfield  {author} {\bibinfo {author} {\bibfnamefont {C.}~\bibnamefont
  {Ryan-Anderson}}, \bibinfo {author} {\bibfnamefont {J.~G.}\ \bibnamefont
  {Bohnet}}, \bibinfo {author} {\bibfnamefont {K.}~\bibnamefont {Lee}},
  \bibinfo {author} {\bibfnamefont {D.}~\bibnamefont {Gresh}}, \bibinfo
  {author} {\bibfnamefont {A.}~\bibnamefont {Hankin}}, \bibinfo {author}
  {\bibfnamefont {J.~P.}\ \bibnamefont {Gaebler}}, \bibinfo {author}
  {\bibfnamefont {D.}~\bibnamefont {Francois}}, \bibinfo {author}
  {\bibfnamefont {A.}~\bibnamefont {Chernoguzov}}, \bibinfo {author}
  {\bibfnamefont {D.}~\bibnamefont {Lucchetti}}, \bibinfo {author}
  {\bibfnamefont {N.~C.}\ \bibnamefont {Brown}}, \bibinfo {author}
  {\bibfnamefont {T.~M.}\ \bibnamefont {Gatterman}}, \bibinfo {author}
  {\bibfnamefont {S.~K.}\ \bibnamefont {Halit}}, \bibinfo {author}
  {\bibfnamefont {K.}~\bibnamefont {Gilmore}}, \bibinfo {author} {\bibfnamefont
  {J.~A.}\ \bibnamefont {Gerber}}, \bibinfo {author} {\bibfnamefont
  {B.}~\bibnamefont {Neyenhuis}}, \bibinfo {author} {\bibfnamefont
  {D.}~\bibnamefont {Hayes}},  \emph {et~al.},\ }\href {\doibase
  10.1103/PhysRevX.11.041058} {\bibfield  {journal} {\bibinfo  {journal} {Phys.
  Rev. X}\ }\textbf {\bibinfo {volume} {11}},\ \bibinfo {pages} {041058}
  (\bibinfo {year} {2021})}\BibitemShut {NoStop}%
\bibitem [{\citenamefont {Suzuki}(1976)}]{trotter}%
  \BibitemOpen
  \bibfield  {author} {\bibinfo {author} {\bibfnamefont {M.}~\bibnamefont
  {Suzuki}},\ }\href {https://doi.org/10.1007/BF01609348} {\bibfield  {journal}
  {\bibinfo  {journal} {Communications in Mathematical Physics}\ }\textbf
  {\bibinfo {volume} {51}},\ \bibinfo {pages} {183} (\bibinfo {year}
  {1976})}\BibitemShut {NoStop}%
\bibitem [{\citenamefont {Arute}\ \emph {et~al.}(2019)\citenamefont {Arute},
  \citenamefont {Arya}, \citenamefont {Babbush}, \citenamefont {Bacon},
  \citenamefont {Bardin}, \citenamefont {Barends}, \citenamefont {Biswas},
  \citenamefont {Boixo}, \citenamefont {Brandao}, \citenamefont {Buell} \emph
  {et~al.}}]{arute2019quantum}%
  \BibitemOpen
  \bibfield  {author} {\bibinfo {author} {\bibfnamefont {F.}~\bibnamefont
  {Arute}}, \bibinfo {author} {\bibfnamefont {K.}~\bibnamefont {Arya}},
  \bibinfo {author} {\bibfnamefont {R.}~\bibnamefont {Babbush}}, \bibinfo
  {author} {\bibfnamefont {D.}~\bibnamefont {Bacon}}, \bibinfo {author}
  {\bibfnamefont {J.~C.}\ \bibnamefont {Bardin}}, \bibinfo {author}
  {\bibfnamefont {R.}~\bibnamefont {Barends}}, \bibinfo {author} {\bibfnamefont
  {R.}~\bibnamefont {Biswas}}, \bibinfo {author} {\bibfnamefont
  {S.}~\bibnamefont {Boixo}}, \bibinfo {author} {\bibfnamefont {F.~G.}\
  \bibnamefont {Brandao}}, \bibinfo {author} {\bibfnamefont {D.~A.}\
  \bibnamefont {Buell}},  \emph {et~al.},\ }\href
  {https://www.nature.com/articles/s41586%20019%201666%205} {\bibfield
  {journal} {\bibinfo  {journal} {Nature}\ }\textbf {\bibinfo {volume} {574}},\
  \bibinfo {pages} {505} (\bibinfo {year} {2019})}\BibitemShut {NoStop}%
\bibitem [{\citenamefont {Zhong}\ \emph {et~al.}(2020)\citenamefont {Zhong},
  \citenamefont {Wang}, \citenamefont {Deng}, \citenamefont {Chen},
  \citenamefont {Peng}, \citenamefont {Luo}, \citenamefont {Qin}, \citenamefont
  {Wu}, \citenamefont {Ding}, \citenamefont {Hu}, \citenamefont {Hu},
  \citenamefont {Yang}, \citenamefont {Zhang}, \citenamefont {Li},
  \citenamefont {Li} \emph {et~al.}}]{doi:10.1126/science.abe8770}%
  \BibitemOpen
  \bibfield  {author} {\bibinfo {author} {\bibfnamefont {H.-S.}\ \bibnamefont
  {Zhong}}, \bibinfo {author} {\bibfnamefont {H.}~\bibnamefont {Wang}},
  \bibinfo {author} {\bibfnamefont {Y.-H.}\ \bibnamefont {Deng}}, \bibinfo
  {author} {\bibfnamefont {M.-C.}\ \bibnamefont {Chen}}, \bibinfo {author}
  {\bibfnamefont {L.-C.}\ \bibnamefont {Peng}}, \bibinfo {author}
  {\bibfnamefont {Y.-H.}\ \bibnamefont {Luo}}, \bibinfo {author} {\bibfnamefont
  {J.}~\bibnamefont {Qin}}, \bibinfo {author} {\bibfnamefont {D.}~\bibnamefont
  {Wu}}, \bibinfo {author} {\bibfnamefont {X.}~\bibnamefont {Ding}}, \bibinfo
  {author} {\bibfnamefont {Y.}~\bibnamefont {Hu}}, \bibinfo {author}
  {\bibfnamefont {P.}~\bibnamefont {Hu}}, \bibinfo {author} {\bibfnamefont
  {X.-Y.}\ \bibnamefont {Yang}}, \bibinfo {author} {\bibfnamefont {W.-J.}\
  \bibnamefont {Zhang}}, \bibinfo {author} {\bibfnamefont {H.}~\bibnamefont
  {Li}}, \bibinfo {author} {\bibfnamefont {Y.}~\bibnamefont {Li}},  \emph
  {et~al.},\ }\href {\doibase 10.1126/science.abe8770} {\bibfield  {journal}
  {\bibinfo  {journal} {Science}\ }\textbf {\bibinfo {volume} {370}},\ \bibinfo
  {pages} {1460} (\bibinfo {year} {2020})},\ \Eprint
  {http://arxiv.org/abs/https://doi.org/10.1126/science.abe8770}
  {https://doi.org/10.1126/science.abe8770} \BibitemShut {NoStop}%
\bibitem [{\citenamefont {Kim}\ \emph {et~al.}(2023)\citenamefont {Kim},
  \citenamefont {Eddins}, \citenamefont {Anand}, \citenamefont {Wei},
  \citenamefont {Van Den~Berg}, \citenamefont {Rosenblatt}, \citenamefont
  {Nayfeh}, \citenamefont {Wu}, \citenamefont {Zaletel}, \citenamefont
  {Temme},\ and\ \citenamefont {Kandala}}]{kim2023evidence}%
  \BibitemOpen
  \bibfield  {author} {\bibinfo {author} {\bibfnamefont {Y.}~\bibnamefont
  {Kim}}, \bibinfo {author} {\bibfnamefont {A.}~\bibnamefont {Eddins}},
  \bibinfo {author} {\bibfnamefont {S.}~\bibnamefont {Anand}}, \bibinfo
  {author} {\bibfnamefont {K.~X.}\ \bibnamefont {Wei}}, \bibinfo {author}
  {\bibfnamefont {E.}~\bibnamefont {Van Den~Berg}}, \bibinfo {author}
  {\bibfnamefont {S.}~\bibnamefont {Rosenblatt}}, \bibinfo {author}
  {\bibfnamefont {H.}~\bibnamefont {Nayfeh}}, \bibinfo {author} {\bibfnamefont
  {Y.}~\bibnamefont {Wu}}, \bibinfo {author} {\bibfnamefont {M.}~\bibnamefont
  {Zaletel}}, \bibinfo {author} {\bibfnamefont {K.}~\bibnamefont {Temme}}, \
  and\ \bibinfo {author} {\bibfnamefont {A.}~\bibnamefont {Kandala}},\ }\href
  {https://www.nature.com/articles/s41586-023-06096-3} {\bibfield  {journal}
  {\bibinfo  {journal} {Nature}\ }\textbf {\bibinfo {volume} {618}},\ \bibinfo
  {pages} {500} (\bibinfo {year} {2023})}\BibitemShut {NoStop}%
\bibitem [{\citenamefont {Patra}\ \emph {et~al.}(2023)\citenamefont {Patra},
  \citenamefont {Jahromi}, \citenamefont {Singh},\ and\ \citenamefont
  {Orus}}]{patra2023efficient}%
  \BibitemOpen
  \bibfield  {author} {\bibinfo {author} {\bibfnamefont {S.}~\bibnamefont
  {Patra}}, \bibinfo {author} {\bibfnamefont {S.~S.}\ \bibnamefont {Jahromi}},
  \bibinfo {author} {\bibfnamefont {S.}~\bibnamefont {Singh}}, \ and\ \bibinfo
  {author} {\bibfnamefont {R.}~\bibnamefont {Orus}},\ }\href@noop {} {\
  (\bibinfo {year} {2023})},\ \Eprint {http://arxiv.org/abs/2309.15642}
  {arXiv:2309.15642 [quant-ph]} \BibitemShut {NoStop}%
\bibitem [{\citenamefont {Tindall}\ \emph {et~al.}(2023)\citenamefont
  {Tindall}, \citenamefont {Fishman}, \citenamefont {Stoudenmire},\ and\
  \citenamefont {Sels}}]{tindall2023efficient}%
  \BibitemOpen
  \bibfield  {author} {\bibinfo {author} {\bibfnamefont {J.}~\bibnamefont
  {Tindall}}, \bibinfo {author} {\bibfnamefont {M.}~\bibnamefont {Fishman}},
  \bibinfo {author} {\bibfnamefont {M.}~\bibnamefont {Stoudenmire}}, \ and\
  \bibinfo {author} {\bibfnamefont {D.}~\bibnamefont {Sels}},\ }\href@noop {}
  {\  (\bibinfo {year} {2023})},\ \Eprint {http://arxiv.org/abs/2306.14887}
  {arXiv:2306.14887 [quant-ph]} \BibitemShut {NoStop}%
\bibitem [{\citenamefont {Liao}\ \emph {et~al.}(2023)\citenamefont {Liao},
  \citenamefont {Wang}, \citenamefont {Zhou}, \citenamefont {Zhang},\ and\
  \citenamefont {Xiang}}]{liao2023simulation}%
  \BibitemOpen
  \bibfield  {author} {\bibinfo {author} {\bibfnamefont {H.-J.}\ \bibnamefont
  {Liao}}, \bibinfo {author} {\bibfnamefont {K.}~\bibnamefont {Wang}}, \bibinfo
  {author} {\bibfnamefont {Z.-S.}\ \bibnamefont {Zhou}}, \bibinfo {author}
  {\bibfnamefont {P.}~\bibnamefont {Zhang}}, \ and\ \bibinfo {author}
  {\bibfnamefont {T.}~\bibnamefont {Xiang}},\ }\href@noop {} {\  (\bibinfo
  {year} {2023})},\ \Eprint {http://arxiv.org/abs/2308.03082} {arXiv:2308.03082
  [quant-ph]} \BibitemShut {NoStop}%
\bibitem [{\citenamefont {Elfving}\ \emph {et~al.}(2020)\citenamefont
  {Elfving}, \citenamefont {Broer}, \citenamefont {Webber}, \citenamefont
  {Gavartin}, \citenamefont {Halls}, \citenamefont {Lorton},\ and\
  \citenamefont {Bochevarov}}]{elfving2020quantum}%
  \BibitemOpen
  \bibfield  {author} {\bibinfo {author} {\bibfnamefont {V.~E.}\ \bibnamefont
  {Elfving}}, \bibinfo {author} {\bibfnamefont {B.~W.}\ \bibnamefont {Broer}},
  \bibinfo {author} {\bibfnamefont {M.}~\bibnamefont {Webber}}, \bibinfo
  {author} {\bibfnamefont {J.}~\bibnamefont {Gavartin}}, \bibinfo {author}
  {\bibfnamefont {M.~D.}\ \bibnamefont {Halls}}, \bibinfo {author}
  {\bibfnamefont {K.~P.}\ \bibnamefont {Lorton}}, \ and\ \bibinfo {author}
  {\bibfnamefont {A.}~\bibnamefont {Bochevarov}},\ }\href@noop {} {\  (\bibinfo
  {year} {2020})},\ \Eprint {http://arxiv.org/abs/2009.12472} {arXiv:2009.12472
  [quant-ph]} \BibitemShut {NoStop}%
\bibitem [{\citenamefont {Daley}\ \emph {et~al.}(2022)\citenamefont {Daley},
  \citenamefont {Bloch}, \citenamefont {Kokail}, \citenamefont {Flannigan},
  \citenamefont {Pearson}, \citenamefont {Troyer},\ and\ \citenamefont
  {Zoller}}]{daley2022practical}%
  \BibitemOpen
  \bibfield  {author} {\bibinfo {author} {\bibfnamefont {A.~J.}\ \bibnamefont
  {Daley}}, \bibinfo {author} {\bibfnamefont {I.}~\bibnamefont {Bloch}},
  \bibinfo {author} {\bibfnamefont {C.}~\bibnamefont {Kokail}}, \bibinfo
  {author} {\bibfnamefont {S.}~\bibnamefont {Flannigan}}, \bibinfo {author}
  {\bibfnamefont {N.}~\bibnamefont {Pearson}}, \bibinfo {author} {\bibfnamefont
  {M.}~\bibnamefont {Troyer}}, \ and\ \bibinfo {author} {\bibfnamefont
  {P.}~\bibnamefont {Zoller}},\ }\href
  {https://doi.org/10.1038/s41586-022-04940-6} {\bibfield  {journal} {\bibinfo
  {journal} {Nature}\ }\textbf {\bibinfo {volume} {607}},\ \bibinfo {pages}
  {667} (\bibinfo {year} {2022})}\BibitemShut {NoStop}%
\bibitem [{\citenamefont {Or{\'u}s}\ \emph {et~al.}(2019)\citenamefont
  {Or{\'u}s}, \citenamefont {Mugel},\ and\ \citenamefont
  {Lizaso}}]{orus2019quantum}%
  \BibitemOpen
  \bibfield  {author} {\bibinfo {author} {\bibfnamefont {R.}~\bibnamefont
  {Or{\'u}s}}, \bibinfo {author} {\bibfnamefont {S.}~\bibnamefont {Mugel}}, \
  and\ \bibinfo {author} {\bibfnamefont {E.}~\bibnamefont {Lizaso}},\ }\href
  {https://doi.org/10.1016/j.revip.2019.100028} {\bibfield  {journal} {\bibinfo
   {journal} {Reviews in Physics}\ }\textbf {\bibinfo {volume} {4}},\ \bibinfo
  {pages} {100028} (\bibinfo {year} {2019})}\BibitemShut {NoStop}%
\bibitem [{\citenamefont {Cao}\ \emph {et~al.}(2018)\citenamefont {Cao},
  \citenamefont {Romero},\ and\ \citenamefont
  {Aspuru-Guzik}}]{cao2018potential}%
  \BibitemOpen
  \bibfield  {author} {\bibinfo {author} {\bibfnamefont {Y.}~\bibnamefont
  {Cao}}, \bibinfo {author} {\bibfnamefont {J.}~\bibnamefont {Romero}}, \ and\
  \bibinfo {author} {\bibfnamefont {A.}~\bibnamefont {Aspuru-Guzik}},\ }\href
  {https://ieeexplore.ieee.org/document/8585034} {\bibfield  {journal}
  {\bibinfo  {journal} {IBM Journal of Research and Development}\ }\textbf
  {\bibinfo {volume} {62}},\ \bibinfo {pages} {6} (\bibinfo {year}
  {2018})}\BibitemShut {NoStop}%
\bibitem [{\citenamefont {Schuld}\ and\ \citenamefont
  {Petruccione}(2021)}]{schuld2021machine}%
  \BibitemOpen
  \bibfield  {author} {\bibinfo {author} {\bibfnamefont {M.}~\bibnamefont
  {Schuld}}\ and\ \bibinfo {author} {\bibfnamefont {F.}~\bibnamefont
  {Petruccione}},\ }\href {https://doi.org/10.1007/978-3-030-83098-4} {\emph
  {\bibinfo {title} {Machine learning with quantum computers}}}\ (\bibinfo
  {publisher} {Springer},\ \bibinfo {year} {2021})\BibitemShut {NoStop}%
\bibitem [{\citenamefont {Bharti}\ \emph {et~al.}(2022)\citenamefont {Bharti},
  \citenamefont {Cervera-Lierta}, \citenamefont {Kyaw}, \citenamefont {Haug},
  \citenamefont {Alperin-Lea}, \citenamefont {Anand}, \citenamefont {Degroote},
  \citenamefont {Heimonen}, \citenamefont {Kottmann}, \citenamefont {Menke},
  \citenamefont {Mok}, \citenamefont {Sim}, \citenamefont {Kwek},\ and\
  \citenamefont {Aspuru-Guzik}}]{Bharti_2022}%
  \BibitemOpen
  \bibfield  {author} {\bibinfo {author} {\bibfnamefont {K.}~\bibnamefont
  {Bharti}}, \bibinfo {author} {\bibfnamefont {A.}~\bibnamefont
  {Cervera-Lierta}}, \bibinfo {author} {\bibfnamefont {T.~H.}\ \bibnamefont
  {Kyaw}}, \bibinfo {author} {\bibfnamefont {T.}~\bibnamefont {Haug}}, \bibinfo
  {author} {\bibfnamefont {S.}~\bibnamefont {Alperin-Lea}}, \bibinfo {author}
  {\bibfnamefont {A.}~\bibnamefont {Anand}}, \bibinfo {author} {\bibfnamefont
  {M.}~\bibnamefont {Degroote}}, \bibinfo {author} {\bibfnamefont
  {H.}~\bibnamefont {Heimonen}}, \bibinfo {author} {\bibfnamefont {J.~S.}\
  \bibnamefont {Kottmann}}, \bibinfo {author} {\bibfnamefont {T.}~\bibnamefont
  {Menke}}, \bibinfo {author} {\bibfnamefont {W.-K.}\ \bibnamefont {Mok}},
  \bibinfo {author} {\bibfnamefont {S.}~\bibnamefont {Sim}}, \bibinfo {author}
  {\bibfnamefont {L.-C.}\ \bibnamefont {Kwek}}, \ and\ \bibinfo {author}
  {\bibfnamefont {A.}~\bibnamefont {Aspuru-Guzik}},\ }\href {\doibase
  10.1103/revmodphys.94.015004} {\bibfield  {journal} {\bibinfo  {journal}
  {Reviews of Modern Physics}\ }\textbf {\bibinfo {volume} {94}} (\bibinfo
  {year} {2022}),\ 10.1103/revmodphys.94.015004}\BibitemShut {NoStop}%
\bibitem [{\citenamefont {Cerezo}\ \emph {et~al.}(2021)\citenamefont {Cerezo},
  \citenamefont {Arrasmith}, \citenamefont {Babbush}, \citenamefont {Benjamin},
  \citenamefont {Endo}, \citenamefont {Fujii}, \citenamefont {McClean},
  \citenamefont {Mitarai}, \citenamefont {Yuan}, \citenamefont {Cincio},\ and\
  \citenamefont {Coles}}]{Cerezo_2021}%
  \BibitemOpen
  \bibfield  {author} {\bibinfo {author} {\bibfnamefont {M.}~\bibnamefont
  {Cerezo}}, \bibinfo {author} {\bibfnamefont {A.}~\bibnamefont {Arrasmith}},
  \bibinfo {author} {\bibfnamefont {R.}~\bibnamefont {Babbush}}, \bibinfo
  {author} {\bibfnamefont {S.~C.}\ \bibnamefont {Benjamin}}, \bibinfo {author}
  {\bibfnamefont {S.}~\bibnamefont {Endo}}, \bibinfo {author} {\bibfnamefont
  {K.}~\bibnamefont {Fujii}}, \bibinfo {author} {\bibfnamefont {J.~R.}\
  \bibnamefont {McClean}}, \bibinfo {author} {\bibfnamefont {K.}~\bibnamefont
  {Mitarai}}, \bibinfo {author} {\bibfnamefont {X.}~\bibnamefont {Yuan}},
  \bibinfo {author} {\bibfnamefont {L.}~\bibnamefont {Cincio}}, \ and\ \bibinfo
  {author} {\bibfnamefont {P.~J.}\ \bibnamefont {Coles}},\ }\href {\doibase
  10.1038/s42254-021-00348-9} {\bibfield  {journal} {\bibinfo  {journal}
  {Nature Reviews Physics}\ }\textbf {\bibinfo {volume} {3}},\ \bibinfo {pages}
  {625–644} (\bibinfo {year} {2021})}\BibitemShut {NoStop}%
\bibitem [{\citenamefont {Dawid}\ \emph {et~al.}(2023)\citenamefont {Dawid},
  \citenamefont {Arnold}, \citenamefont {Requena}, \citenamefont {Gresch},
  \citenamefont {Płodzień}, \citenamefont {Donatella}, \citenamefont
  {Nicoli}, \citenamefont {Stornati}, \citenamefont {Koch}, \citenamefont
  {Büttner}, \citenamefont {Okuła}, \citenamefont {Muñoz-Gil}, \citenamefont
  {Vargas-Hernández}, \citenamefont {Cervera-Lierta}, \citenamefont
  {Carrasquilla} \emph {et~al.}}]{dawid2023modern}%
  \BibitemOpen
  \bibfield  {author} {\bibinfo {author} {\bibfnamefont {A.}~\bibnamefont
  {Dawid}}, \bibinfo {author} {\bibfnamefont {J.}~\bibnamefont {Arnold}},
  \bibinfo {author} {\bibfnamefont {B.}~\bibnamefont {Requena}}, \bibinfo
  {author} {\bibfnamefont {A.}~\bibnamefont {Gresch}}, \bibinfo {author}
  {\bibfnamefont {M.}~\bibnamefont {Płodzień}}, \bibinfo {author}
  {\bibfnamefont {K.}~\bibnamefont {Donatella}}, \bibinfo {author}
  {\bibfnamefont {K.~A.}\ \bibnamefont {Nicoli}}, \bibinfo {author}
  {\bibfnamefont {P.}~\bibnamefont {Stornati}}, \bibinfo {author}
  {\bibfnamefont {R.}~\bibnamefont {Koch}}, \bibinfo {author} {\bibfnamefont
  {M.}~\bibnamefont {Büttner}}, \bibinfo {author} {\bibfnamefont
  {R.}~\bibnamefont {Okuła}}, \bibinfo {author} {\bibfnamefont
  {G.}~\bibnamefont {Muñoz-Gil}}, \bibinfo {author} {\bibfnamefont {R.~A.}\
  \bibnamefont {Vargas-Hernández}}, \bibinfo {author} {\bibfnamefont
  {A.}~\bibnamefont {Cervera-Lierta}}, \bibinfo {author} {\bibfnamefont
  {J.}~\bibnamefont {Carrasquilla}},  \emph {et~al.},\ }\href@noop {} {\
  (\bibinfo {year} {2023})},\ \Eprint {http://arxiv.org/abs/2204.04198}
  {arXiv:2204.04198 [quant-ph]} \BibitemShut {NoStop}%
\bibitem [{\citenamefont {Zhang}\ \emph {et~al.}(2017)\citenamefont {Zhang},
  \citenamefont {Pagano}, \citenamefont {Hess}, \citenamefont {Kyprianidis},
  \citenamefont {Becker}, \citenamefont {Kaplan}, \citenamefont {Gorshkov},
  \citenamefont {Gong},\ and\ \citenamefont {Monroe}}]{Zhang_2017}%
  \BibitemOpen
  \bibfield  {author} {\bibinfo {author} {\bibfnamefont {J.}~\bibnamefont
  {Zhang}}, \bibinfo {author} {\bibfnamefont {G.}~\bibnamefont {Pagano}},
  \bibinfo {author} {\bibfnamefont {P.~W.}\ \bibnamefont {Hess}}, \bibinfo
  {author} {\bibfnamefont {A.}~\bibnamefont {Kyprianidis}}, \bibinfo {author}
  {\bibfnamefont {P.}~\bibnamefont {Becker}}, \bibinfo {author} {\bibfnamefont
  {H.}~\bibnamefont {Kaplan}}, \bibinfo {author} {\bibfnamefont {A.~V.}\
  \bibnamefont {Gorshkov}}, \bibinfo {author} {\bibfnamefont {Z.-X.}\
  \bibnamefont {Gong}}, \ and\ \bibinfo {author} {\bibfnamefont
  {C.}~\bibnamefont {Monroe}},\ }\href {\doibase 10.1038/nature24654}
  {\bibfield  {journal} {\bibinfo  {journal} {Nature}\ }\textbf {\bibinfo
  {volume} {551}},\ \bibinfo {pages} {601} (\bibinfo {year}
  {2017})}\BibitemShut {NoStop}%
\bibitem [{\citenamefont {yoon Choi}\ \emph {et~al.}(2016)\citenamefont {yoon
  Choi}, \citenamefont {Hild}, \citenamefont {Zeiher}, \citenamefont
  {Schau{\ss}}, \citenamefont {Rubio-Abadal}, \citenamefont {Yefsah},
  \citenamefont {Khemani}, \citenamefont {Huse}, \citenamefont {Bloch},\ and\
  \citenamefont {Gross}}]{Choi_2016}%
  \BibitemOpen
  \bibfield  {author} {\bibinfo {author} {\bibfnamefont {J.}~\bibnamefont {yoon
  Choi}}, \bibinfo {author} {\bibfnamefont {S.}~\bibnamefont {Hild}}, \bibinfo
  {author} {\bibfnamefont {J.}~\bibnamefont {Zeiher}}, \bibinfo {author}
  {\bibfnamefont {P.}~\bibnamefont {Schau{\ss}}}, \bibinfo {author}
  {\bibfnamefont {A.}~\bibnamefont {Rubio-Abadal}}, \bibinfo {author}
  {\bibfnamefont {T.}~\bibnamefont {Yefsah}}, \bibinfo {author} {\bibfnamefont
  {V.}~\bibnamefont {Khemani}}, \bibinfo {author} {\bibfnamefont {D.~A.}\
  \bibnamefont {Huse}}, \bibinfo {author} {\bibfnamefont {I.}~\bibnamefont
  {Bloch}}, \ and\ \bibinfo {author} {\bibfnamefont {C.}~\bibnamefont
  {Gross}},\ }\href {\doibase 10.1126/science.aaf8834} {\bibfield  {journal}
  {\bibinfo  {journal} {Science}\ }\textbf {\bibinfo {volume} {352}},\ \bibinfo
  {pages} {1547} (\bibinfo {year} {2016})}\BibitemShut {NoStop}%
\bibitem [{\citenamefont {Tillmann}\ \emph {et~al.}(2013)\citenamefont
  {Tillmann}, \citenamefont {Daki{\'{c}}}, \citenamefont {Heilmann},
  \citenamefont {Nolte}, \citenamefont {Szameit},\ and\ \citenamefont
  {Walther}}]{Tillmann_2013}%
  \BibitemOpen
  \bibfield  {author} {\bibinfo {author} {\bibfnamefont {M.}~\bibnamefont
  {Tillmann}}, \bibinfo {author} {\bibfnamefont {B.}~\bibnamefont
  {Daki{\'{c}}}}, \bibinfo {author} {\bibfnamefont {R.}~\bibnamefont
  {Heilmann}}, \bibinfo {author} {\bibfnamefont {S.}~\bibnamefont {Nolte}},
  \bibinfo {author} {\bibfnamefont {A.}~\bibnamefont {Szameit}}, \ and\
  \bibinfo {author} {\bibfnamefont {P.}~\bibnamefont {Walther}},\ }\href
  {\doibase 10.1038/nphoton.2013.102} {\bibfield  {journal} {\bibinfo
  {journal} {Nature Photonics}\ }\textbf {\bibinfo {volume} {7}},\ \bibinfo
  {pages} {540} (\bibinfo {year} {2013})}\BibitemShut {NoStop}%
\bibitem [{\citenamefont {Briegel}\ \emph {et~al.}(2000)\citenamefont
  {Briegel}, \citenamefont {Calarco}, \citenamefont {Jaksch}, \citenamefont
  {Cirac},\ and\ \citenamefont {Zoller}}]{briegel2000quantum}%
  \BibitemOpen
  \bibfield  {author} {\bibinfo {author} {\bibfnamefont {H.-J.}\ \bibnamefont
  {Briegel}}, \bibinfo {author} {\bibfnamefont {T.}~\bibnamefont {Calarco}},
  \bibinfo {author} {\bibfnamefont {D.}~\bibnamefont {Jaksch}}, \bibinfo
  {author} {\bibfnamefont {J.~I.}\ \bibnamefont {Cirac}}, \ and\ \bibinfo
  {author} {\bibfnamefont {P.}~\bibnamefont {Zoller}},\ }\href
  {https://doi.org/10.1080/09500340008244052} {\bibfield  {journal} {\bibinfo
  {journal} {Journal of modern optics}\ }\textbf {\bibinfo {volume} {47}},\
  \bibinfo {pages} {415} (\bibinfo {year} {2000})}\BibitemShut {NoStop}%
\bibitem [{\citenamefont {Bloch}\ \emph {et~al.}(2012)\citenamefont {Bloch},
  \citenamefont {Dalibard},\ and\ \citenamefont
  {Nascimbene}}]{bloch2012quantum}%
  \BibitemOpen
  \bibfield  {author} {\bibinfo {author} {\bibfnamefont {I.}~\bibnamefont
  {Bloch}}, \bibinfo {author} {\bibfnamefont {J.}~\bibnamefont {Dalibard}}, \
  and\ \bibinfo {author} {\bibfnamefont {S.}~\bibnamefont {Nascimbene}},\
  }\href@noop {} {\bibfield  {journal} {\bibinfo  {journal} {Nature Physics}\
  }\textbf {\bibinfo {volume} {8}},\ \bibinfo {pages} {267} (\bibinfo {year}
  {2012})}\BibitemShut {NoStop}%
\bibitem [{\citenamefont {Preskill}(2018)}]{Preskill_2018}%
  \BibitemOpen
  \bibfield  {author} {\bibinfo {author} {\bibfnamefont {J.}~\bibnamefont
  {Preskill}},\ }\href {\doibase 10.22331/q-2018-08-06-79} {\bibfield
  {journal} {\bibinfo  {journal} {Quantum}\ }\textbf {\bibinfo {volume} {2}},\
  \bibinfo {pages} {79} (\bibinfo {year} {2018})}\BibitemShut {NoStop}%
\bibitem [{\citenamefont {Scholl}\ \emph {et~al.}(2021)\citenamefont {Scholl},
  \citenamefont {Schuler}, \citenamefont {Williams}, \citenamefont
  {Eberharter}, \citenamefont {Barredo}, \citenamefont {Schymik}, \citenamefont
  {Lienhard}, \citenamefont {Henry}, \citenamefont {Lang}, \citenamefont
  {Lahaye}, \citenamefont {Läuchli},\ and\ \citenamefont
  {Browaeys}}]{Scholl_2021}%
  \BibitemOpen
  \bibfield  {author} {\bibinfo {author} {\bibfnamefont {P.}~\bibnamefont
  {Scholl}}, \bibinfo {author} {\bibfnamefont {M.}~\bibnamefont {Schuler}},
  \bibinfo {author} {\bibfnamefont {H.~J.}\ \bibnamefont {Williams}}, \bibinfo
  {author} {\bibfnamefont {A.~A.}\ \bibnamefont {Eberharter}}, \bibinfo
  {author} {\bibfnamefont {D.}~\bibnamefont {Barredo}}, \bibinfo {author}
  {\bibfnamefont {K.-N.}\ \bibnamefont {Schymik}}, \bibinfo {author}
  {\bibfnamefont {V.}~\bibnamefont {Lienhard}}, \bibinfo {author}
  {\bibfnamefont {L.-P.}\ \bibnamefont {Henry}}, \bibinfo {author}
  {\bibfnamefont {T.~C.}\ \bibnamefont {Lang}}, \bibinfo {author}
  {\bibfnamefont {T.}~\bibnamefont {Lahaye}}, \bibinfo {author} {\bibfnamefont
  {A.~M.}\ \bibnamefont {Läuchli}}, \ and\ \bibinfo {author} {\bibfnamefont
  {A.}~\bibnamefont {Browaeys}},\ }\href {\doibase 10.1038/s41586-021-03585-1}
  {\bibfield  {journal} {\bibinfo  {journal} {Nature}\ }\textbf {\bibinfo
  {volume} {595}},\ \bibinfo {pages} {233} (\bibinfo {year}
  {2021})}\BibitemShut {NoStop}%
\bibitem [{\citenamefont {Shaw}\ \emph {et~al.}(2023)\citenamefont {Shaw},
  \citenamefont {Chen}, \citenamefont {Choi}, \citenamefont {Mark},
  \citenamefont {Scholl}, \citenamefont {Finkelstein}, \citenamefont {Elben},
  \citenamefont {Choi},\ and\ \citenamefont {Endres}}]{shaw2023benchmarking}%
  \BibitemOpen
  \bibfield  {author} {\bibinfo {author} {\bibfnamefont {A.~L.}\ \bibnamefont
  {Shaw}}, \bibinfo {author} {\bibfnamefont {Z.}~\bibnamefont {Chen}}, \bibinfo
  {author} {\bibfnamefont {J.}~\bibnamefont {Choi}}, \bibinfo {author}
  {\bibfnamefont {D.~K.}\ \bibnamefont {Mark}}, \bibinfo {author}
  {\bibfnamefont {P.}~\bibnamefont {Scholl}}, \bibinfo {author} {\bibfnamefont
  {R.}~\bibnamefont {Finkelstein}}, \bibinfo {author} {\bibfnamefont
  {A.}~\bibnamefont {Elben}}, \bibinfo {author} {\bibfnamefont
  {S.}~\bibnamefont {Choi}}, \ and\ \bibinfo {author} {\bibfnamefont
  {M.}~\bibnamefont {Endres}},\ }\href@noop {} {\  (\bibinfo {year} {2023})},\
  \Eprint {http://arxiv.org/abs/2308.07914} {arXiv:2308.07914 [quant-ph]}
  \BibitemShut {NoStop}%
\bibitem [{\citenamefont {Jones}(2011)}]{JONES201191}%
  \BibitemOpen
  \bibfield  {author} {\bibinfo {author} {\bibfnamefont {J.~A.}\ \bibnamefont
  {Jones}},\ }\href {\doibase https://doi.org/10.1016/j.pnmrs.2010.11.001}
  {\bibfield  {journal} {\bibinfo  {journal} {Progress in Nuclear Magnetic
  Resonance Spectroscopy}\ }\textbf {\bibinfo {volume} {59}},\ \bibinfo {pages}
  {91} (\bibinfo {year} {2011})}\BibitemShut {NoStop}%
\bibitem [{\citenamefont {Lamata}\ \emph {et~al.}(2018)\citenamefont {Lamata},
  \citenamefont {Parra-Rodriguez}, \citenamefont {Sanz},\ and\ \citenamefont
  {Solano}}]{Lamata_2018}%
  \BibitemOpen
  \bibfield  {author} {\bibinfo {author} {\bibfnamefont {L.}~\bibnamefont
  {Lamata}}, \bibinfo {author} {\bibfnamefont {A.}~\bibnamefont
  {Parra-Rodriguez}}, \bibinfo {author} {\bibfnamefont {M.}~\bibnamefont
  {Sanz}}, \ and\ \bibinfo {author} {\bibfnamefont {E.}~\bibnamefont
  {Solano}},\ }\href {\doibase 10.1080/23746149.2018.1457981} {\bibfield
  {journal} {\bibinfo  {journal} {Advances in Physics: X}\ }\textbf {\bibinfo
  {volume} {3}},\ \bibinfo {pages} {1457981} (\bibinfo {year}
  {2018})}\BibitemShut {NoStop}%
\bibitem [{\citenamefont {Gonzalez-Raya}\ \emph {et~al.}(2021)\citenamefont
  {Gonzalez-Raya}, \citenamefont {Asensio-Perea}, \citenamefont {Martin},
  \citenamefont {C\'eleri}, \citenamefont {Sanz}, \citenamefont {Lougovski},\
  and\ \citenamefont {Dumitrescu}}]{PRXQuantum.2.020328}%
  \BibitemOpen
  \bibfield  {author} {\bibinfo {author} {\bibfnamefont {T.}~\bibnamefont
  {Gonzalez-Raya}}, \bibinfo {author} {\bibfnamefont {R.}~\bibnamefont
  {Asensio-Perea}}, \bibinfo {author} {\bibfnamefont {A.}~\bibnamefont
  {Martin}}, \bibinfo {author} {\bibfnamefont {L.~C.}\ \bibnamefont
  {C\'eleri}}, \bibinfo {author} {\bibfnamefont {M.}~\bibnamefont {Sanz}},
  \bibinfo {author} {\bibfnamefont {P.}~\bibnamefont {Lougovski}}, \ and\
  \bibinfo {author} {\bibfnamefont {E.~F.}\ \bibnamefont {Dumitrescu}},\ }\href
  {\doibase 10.1103/PRXQuantum.2.020328} {\bibfield  {journal} {\bibinfo
  {journal} {PRX Quantum}\ }\textbf {\bibinfo {volume} {2}},\ \bibinfo {pages}
  {020328} (\bibinfo {year} {2021})}\BibitemShut {NoStop}%
\bibitem [{\citenamefont {Galicia}\ \emph {et~al.}(2020)\citenamefont
  {Galicia}, \citenamefont {Ramon}, \citenamefont {Solano},\ and\ \citenamefont
  {Sanz}}]{PhysRevResearch.2.033103}%
  \BibitemOpen
  \bibfield  {author} {\bibinfo {author} {\bibfnamefont {A.}~\bibnamefont
  {Galicia}}, \bibinfo {author} {\bibfnamefont {B.}~\bibnamefont {Ramon}},
  \bibinfo {author} {\bibfnamefont {E.}~\bibnamefont {Solano}}, \ and\ \bibinfo
  {author} {\bibfnamefont {M.}~\bibnamefont {Sanz}},\ }\href {\doibase
  10.1103/PhysRevResearch.2.033103} {\bibfield  {journal} {\bibinfo  {journal}
  {Phys. Rev. Res.}\ }\textbf {\bibinfo {volume} {2}},\ \bibinfo {pages}
  {033103} (\bibinfo {year} {2020})}\BibitemShut {NoStop}%
\bibitem [{\citenamefont {Greenaway}\ \emph {et~al.}(2023)\citenamefont
  {Greenaway}, \citenamefont {Smith}, \citenamefont {Mintert},\ and\
  \citenamefont {Malz}}]{greenaway2023analogue}%
  \BibitemOpen
  \bibfield  {author} {\bibinfo {author} {\bibfnamefont {S.}~\bibnamefont
  {Greenaway}}, \bibinfo {author} {\bibfnamefont {A.}~\bibnamefont {Smith}},
  \bibinfo {author} {\bibfnamefont {F.}~\bibnamefont {Mintert}}, \ and\
  \bibinfo {author} {\bibfnamefont {D.}~\bibnamefont {Malz}},\ }\href@noop {}
  {\  (\bibinfo {year} {2023})},\ \Eprint {http://arxiv.org/abs/2211.16439}
  {arXiv:2211.16439 [quant-ph]} \BibitemShut {NoStop}%
\bibitem [{\citenamefont {Parra-Rodriguez}\ \emph {et~al.}(2020)\citenamefont
  {Parra-Rodriguez}, \citenamefont {Lougovski}, \citenamefont {Lamata},
  \citenamefont {Solano},\ and\ \citenamefont {Sanz}}]{parra2020digital}%
  \BibitemOpen
  \bibfield  {author} {\bibinfo {author} {\bibfnamefont {A.}~\bibnamefont
  {Parra-Rodriguez}}, \bibinfo {author} {\bibfnamefont {P.}~\bibnamefont
  {Lougovski}}, \bibinfo {author} {\bibfnamefont {L.}~\bibnamefont {Lamata}},
  \bibinfo {author} {\bibfnamefont {E.}~\bibnamefont {Solano}}, \ and\ \bibinfo
  {author} {\bibfnamefont {M.}~\bibnamefont {Sanz}},\ }\href
  {https://doi.org/10.1103/PhysRevA.101.022305} {\bibfield  {journal} {\bibinfo
   {journal} {Physical Review A}\ }\textbf {\bibinfo {volume} {101}},\ \bibinfo
  {pages} {022305} (\bibinfo {year} {2020})}\BibitemShut {NoStop}%
\bibitem [{\citenamefont {de~Andoin}\ \emph {et~al.}(2023)\citenamefont
  {de~Andoin}, \citenamefont {Álvaro Saiz}, \citenamefont {Pérez-Fernández},
  \citenamefont {Lamata}, \citenamefont {Oregi},\ and\ \citenamefont
  {Sanz}}]{garciadeandoin2023digitalanalog}%
  \BibitemOpen
  \bibfield  {author} {\bibinfo {author} {\bibfnamefont {M.~G.}\ \bibnamefont
  {de~Andoin}}, \bibinfo {author} {\bibnamefont {Álvaro Saiz}}, \bibinfo
  {author} {\bibfnamefont {P.}~\bibnamefont {Pérez-Fernández}}, \bibinfo
  {author} {\bibfnamefont {L.}~\bibnamefont {Lamata}}, \bibinfo {author}
  {\bibfnamefont {I.}~\bibnamefont {Oregi}}, \ and\ \bibinfo {author}
  {\bibfnamefont {M.}~\bibnamefont {Sanz}},\ }\href@noop {} {\  (\bibinfo
  {year} {2023})},\ \Eprint {http://arxiv.org/abs/2307.00966} {arXiv:2307.00966
  [quant-ph]} \BibitemShut {NoStop}%
\bibitem [{\citenamefont {Kumar}\ \emph {et~al.}(2023)\citenamefont {Kumar},
  \citenamefont {Hegade}, \citenamefont {Solano}, \citenamefont
  {Albarrán-Arriagada},\ and\ \citenamefont
  {Barrios}}]{kumar2023digitalanalog}%
  \BibitemOpen
  \bibfield  {author} {\bibinfo {author} {\bibfnamefont {S.}~\bibnamefont
  {Kumar}}, \bibinfo {author} {\bibfnamefont {N.~N.}\ \bibnamefont {Hegade}},
  \bibinfo {author} {\bibfnamefont {E.}~\bibnamefont {Solano}}, \bibinfo
  {author} {\bibfnamefont {F.}~\bibnamefont {Albarrán-Arriagada}}, \ and\
  \bibinfo {author} {\bibfnamefont {G.~A.}\ \bibnamefont {Barrios}},\
  }\href@noop {} {\  (\bibinfo {year} {2023})},\ \Eprint
  {http://arxiv.org/abs/2308.12040} {arXiv:2308.12040 [quant-ph]} \BibitemShut
  {NoStop}%
\bibitem [{\citenamefont {Headley}\ \emph {et~al.}(2022)\citenamefont
  {Headley}, \citenamefont {Müller}, \citenamefont {Martin}, \citenamefont
  {Solano}, \citenamefont {Sanz},\ and\ \citenamefont
  {Wilhelm}}]{Headley_2022}%
  \BibitemOpen
  \bibfield  {author} {\bibinfo {author} {\bibfnamefont {D.}~\bibnamefont
  {Headley}}, \bibinfo {author} {\bibfnamefont {T.}~\bibnamefont {Müller}},
  \bibinfo {author} {\bibfnamefont {A.}~\bibnamefont {Martin}}, \bibinfo
  {author} {\bibfnamefont {E.}~\bibnamefont {Solano}}, \bibinfo {author}
  {\bibfnamefont {M.}~\bibnamefont {Sanz}}, \ and\ \bibinfo {author}
  {\bibfnamefont {F.~K.}\ \bibnamefont {Wilhelm}},\ }\href {\doibase
  10.1103/physreva.106.042446} {\bibfield  {journal} {\bibinfo  {journal}
  {Physical Review A}\ }\textbf {\bibinfo {volume} {106}} (\bibinfo {year}
  {2022}),\ 10.1103/physreva.106.042446}\BibitemShut {NoStop}%
\bibitem [{\citenamefont {Martin}\ \emph {et~al.}(2023)\citenamefont {Martin},
  \citenamefont {Ibarrondo},\ and\ \citenamefont
  {Sanz}}]{PhysRevApplied.19.064056}%
  \BibitemOpen
  \bibfield  {author} {\bibinfo {author} {\bibfnamefont {A.}~\bibnamefont
  {Martin}}, \bibinfo {author} {\bibfnamefont {R.}~\bibnamefont {Ibarrondo}}, \
  and\ \bibinfo {author} {\bibfnamefont {M.}~\bibnamefont {Sanz}},\ }\href
  {\doibase 10.1103/PhysRevApplied.19.064056} {\bibfield  {journal} {\bibinfo
  {journal} {Phys. Rev. Appl.}\ }\textbf {\bibinfo {volume} {19}},\ \bibinfo
  {pages} {064056} (\bibinfo {year} {2023})}\BibitemShut {NoStop}%
\bibitem [{\citenamefont {Martin}\ \emph {et~al.}(2020)\citenamefont {Martin},
  \citenamefont {Lamata}, \citenamefont {Solano},\ and\ \citenamefont
  {Sanz}}]{Martin_2020}%
  \BibitemOpen
  \bibfield  {author} {\bibinfo {author} {\bibfnamefont {A.}~\bibnamefont
  {Martin}}, \bibinfo {author} {\bibfnamefont {L.}~\bibnamefont {Lamata}},
  \bibinfo {author} {\bibfnamefont {E.}~\bibnamefont {Solano}}, \ and\ \bibinfo
  {author} {\bibfnamefont {M.}~\bibnamefont {Sanz}},\ }\href {\doibase
  10.1103/physrevresearch.2.013012} {\bibfield  {journal} {\bibinfo  {journal}
  {Physical Review Research}\ }\textbf {\bibinfo {volume} {2}} (\bibinfo {year}
  {2020}),\ 10.1103/physrevresearch.2.013012}\BibitemShut {NoStop}%
\bibitem [{\citenamefont {Egger}\ \emph {et~al.}(2023)\citenamefont {Egger},
  \citenamefont {Capecci}, \citenamefont {Pokharel}, \citenamefont
  {Barkoutsos}, \citenamefont {Fischer}, \citenamefont {Guidoni},\ and\
  \citenamefont {Tavernelli}}]{PhysRevResearch.5.033159}%
  \BibitemOpen
  \bibfield  {author} {\bibinfo {author} {\bibfnamefont {D.~J.}\ \bibnamefont
  {Egger}}, \bibinfo {author} {\bibfnamefont {C.}~\bibnamefont {Capecci}},
  \bibinfo {author} {\bibfnamefont {B.}~\bibnamefont {Pokharel}}, \bibinfo
  {author} {\bibfnamefont {P.~K.}\ \bibnamefont {Barkoutsos}}, \bibinfo
  {author} {\bibfnamefont {L.~E.}\ \bibnamefont {Fischer}}, \bibinfo {author}
  {\bibfnamefont {L.}~\bibnamefont {Guidoni}}, \ and\ \bibinfo {author}
  {\bibfnamefont {I.}~\bibnamefont {Tavernelli}},\ }\href {\doibase
  10.1103/PhysRevResearch.5.033159} {\bibfield  {journal} {\bibinfo  {journal}
  {Phys. Rev. Res.}\ }\textbf {\bibinfo {volume} {5}},\ \bibinfo {pages}
  {033159} (\bibinfo {year} {2023})}\BibitemShut {NoStop}%
\bibitem [{\citenamefont {Watanabe}\ \emph {et~al.}(2024)\citenamefont
  {Watanabe}, \citenamefont {Tabuchi}, \citenamefont {Heya}, \citenamefont
  {Tamate},\ and\ \citenamefont {Nakamura}}]{PhysRevA.109.012616}%
  \BibitemOpen
  \bibfield  {author} {\bibinfo {author} {\bibfnamefont {S.}~\bibnamefont
  {Watanabe}}, \bibinfo {author} {\bibfnamefont {Y.}~\bibnamefont {Tabuchi}},
  \bibinfo {author} {\bibfnamefont {K.}~\bibnamefont {Heya}}, \bibinfo {author}
  {\bibfnamefont {S.}~\bibnamefont {Tamate}}, \ and\ \bibinfo {author}
  {\bibfnamefont {Y.}~\bibnamefont {Nakamura}},\ }\href {\doibase
  10.1103/PhysRevA.109.012616} {\bibfield  {journal} {\bibinfo  {journal}
  {Phys. Rev. A}\ }\textbf {\bibinfo {volume} {109}},\ \bibinfo {pages}
  {012616} (\bibinfo {year} {2024})}\BibitemShut {NoStop}%
\bibitem [{\citenamefont {Cesa}\ and\ \citenamefont
  {Pichler}(2023)}]{PhysRevLett.131.170601}%
  \BibitemOpen
  \bibfield  {author} {\bibinfo {author} {\bibfnamefont {F.}~\bibnamefont
  {Cesa}}\ and\ \bibinfo {author} {\bibfnamefont {H.}~\bibnamefont {Pichler}},\
  }\href {\doibase 10.1103/PhysRevLett.131.170601} {\bibfield  {journal}
  {\bibinfo  {journal} {Phys. Rev. Lett.}\ }\textbf {\bibinfo {volume} {131}},\
  \bibinfo {pages} {170601} (\bibinfo {year} {2023})}\BibitemShut {NoStop}%
\bibitem [{\citenamefont {Jones}\ and\ \citenamefont
  {Knill}(1999)}]{JONES1999322}%
  \BibitemOpen
  \bibfield  {author} {\bibinfo {author} {\bibfnamefont {J.}~\bibnamefont
  {Jones}}\ and\ \bibinfo {author} {\bibfnamefont {E.}~\bibnamefont {Knill}},\
  }\href {\doibase https://doi.org/10.1006/jmre.1999.1890} {\bibfield
  {journal} {\bibinfo  {journal} {Journal of Magnetic Resonance}\ }\textbf
  {\bibinfo {volume} {141}},\ \bibinfo {pages} {322} (\bibinfo {year}
  {1999})}\BibitemShut {NoStop}%
\bibitem [{\citenamefont {Leung}\ \emph {et~al.}(2000)\citenamefont {Leung},
  \citenamefont {Chuang}, \citenamefont {Yamaguchi},\ and\ \citenamefont
  {Yamamoto}}]{PhysRevA.61.042310}%
  \BibitemOpen
  \bibfield  {author} {\bibinfo {author} {\bibfnamefont {D.~W.}\ \bibnamefont
  {Leung}}, \bibinfo {author} {\bibfnamefont {I.~L.}\ \bibnamefont {Chuang}},
  \bibinfo {author} {\bibfnamefont {F.}~\bibnamefont {Yamaguchi}}, \ and\
  \bibinfo {author} {\bibfnamefont {Y.}~\bibnamefont {Yamamoto}},\ }\href
  {\doibase 10.1103/PhysRevA.61.042310} {\bibfield  {journal} {\bibinfo
  {journal} {Phys. Rev. A}\ }\textbf {\bibinfo {volume} {61}},\ \bibinfo
  {pages} {042310} (\bibinfo {year} {2000})}\BibitemShut {NoStop}%
\bibitem [{\citenamefont {Dodd}\ \emph {et~al.}(2002)\citenamefont {Dodd},
  \citenamefont {Nielsen}, \citenamefont {Bremner},\ and\ \citenamefont
  {Thew}}]{PhysRevA.65.040301}%
  \BibitemOpen
  \bibfield  {author} {\bibinfo {author} {\bibfnamefont {J.~L.}\ \bibnamefont
  {Dodd}}, \bibinfo {author} {\bibfnamefont {M.~A.}\ \bibnamefont {Nielsen}},
  \bibinfo {author} {\bibfnamefont {M.~J.}\ \bibnamefont {Bremner}}, \ and\
  \bibinfo {author} {\bibfnamefont {R.~T.}\ \bibnamefont {Thew}},\ }\href
  {\doibase 10.1103/PhysRevA.65.040301} {\bibfield  {journal} {\bibinfo
  {journal} {Phys. Rev. A}\ }\textbf {\bibinfo {volume} {65}},\ \bibinfo
  {pages} {040301} (\bibinfo {year} {2002})}\BibitemShut {NoStop}%
\bibitem [{\citenamefont {Guseynov}\ and\ \citenamefont
  {Pogosov}(2022)}]{Guseynov_2022}%
  \BibitemOpen
  \bibfield  {author} {\bibinfo {author} {\bibfnamefont {N.~M.}\ \bibnamefont
  {Guseynov}}\ and\ \bibinfo {author} {\bibfnamefont {W.~V.}\ \bibnamefont
  {Pogosov}},\ }\href {\doibase 10.1088/1361-648x/ac6927} {\bibfield  {journal}
  {\bibinfo  {journal} {Journal of Physics: Condensed Matter}\ }\textbf
  {\bibinfo {volume} {34}},\ \bibinfo {pages} {285901} (\bibinfo {year}
  {2022})}\BibitemShut {NoStop}%
\bibitem [{\citenamefont {Votto}\ \emph {et~al.}(2024)\citenamefont {Votto},
  \citenamefont {Zeiher},\ and\ \citenamefont {Vermersch}}]{votto2024robust}%
  \BibitemOpen
  \bibfield  {author} {\bibinfo {author} {\bibfnamefont {M.}~\bibnamefont
  {Votto}}, \bibinfo {author} {\bibfnamefont {J.}~\bibnamefont {Zeiher}}, \
  and\ \bibinfo {author} {\bibfnamefont {B.}~\bibnamefont {Vermersch}},\
  }\href@noop {} {\  (\bibinfo {year} {2024})},\ \Eprint
  {http://arxiv.org/abs/2311.10600} {arXiv:2311.10600 [quant-ph]} \BibitemShut
  {NoStop}%
\bibitem [{\citenamefont {Pichler}\ \emph {et~al.}(2018)\citenamefont
  {Pichler}, \citenamefont {Wang}, \citenamefont {Zhou}, \citenamefont {Choi},\
  and\ \citenamefont {Lukin}}]{pichler2018quantum}%
  \BibitemOpen
  \bibfield  {author} {\bibinfo {author} {\bibfnamefont {H.}~\bibnamefont
  {Pichler}}, \bibinfo {author} {\bibfnamefont {S.-T.}\ \bibnamefont {Wang}},
  \bibinfo {author} {\bibfnamefont {L.}~\bibnamefont {Zhou}}, \bibinfo {author}
  {\bibfnamefont {S.}~\bibnamefont {Choi}}, \ and\ \bibinfo {author}
  {\bibfnamefont {M.~D.}\ \bibnamefont {Lukin}},\ }\href@noop {} {\  (\bibinfo
  {year} {2018})},\ \Eprint {http://arxiv.org/abs/1808.10816} {arXiv:1808.10816
  [quant-ph]} \BibitemShut {NoStop}%
\bibitem [{\citenamefont {Wurtz}\ \emph {et~al.}(2022)\citenamefont {Wurtz},
  \citenamefont {Lopes}, \citenamefont {Gemelke}, \citenamefont {Keesling},\
  and\ \citenamefont {Wang}}]{wurtz2022industry}%
  \BibitemOpen
  \bibfield  {author} {\bibinfo {author} {\bibfnamefont {J.}~\bibnamefont
  {Wurtz}}, \bibinfo {author} {\bibfnamefont {P.~L.~S.}\ \bibnamefont {Lopes}},
  \bibinfo {author} {\bibfnamefont {N.}~\bibnamefont {Gemelke}}, \bibinfo
  {author} {\bibfnamefont {A.}~\bibnamefont {Keesling}}, \ and\ \bibinfo
  {author} {\bibfnamefont {S.}~\bibnamefont {Wang}},\ }\href@noop {} {\
  (\bibinfo {year} {2022})},\ \Eprint {http://arxiv.org/abs/2205.08500}
  {arXiv:2205.08500 [quant-ph]} \BibitemShut {NoStop}%
\bibitem [{\citenamefont {Bravo}\ \emph {et~al.}(2022)\citenamefont {Bravo},
  \citenamefont {Najafi}, \citenamefont {Gao},\ and\ \citenamefont
  {Yelin}}]{PRXQuantum.3.030325}%
  \BibitemOpen
  \bibfield  {author} {\bibinfo {author} {\bibfnamefont {R.~A.}\ \bibnamefont
  {Bravo}}, \bibinfo {author} {\bibfnamefont {K.}~\bibnamefont {Najafi}},
  \bibinfo {author} {\bibfnamefont {X.}~\bibnamefont {Gao}}, \ and\ \bibinfo
  {author} {\bibfnamefont {S.~F.}\ \bibnamefont {Yelin}},\ }\href {\doibase
  10.1103/PRXQuantum.3.030325} {\bibfield  {journal} {\bibinfo  {journal} {PRX
  Quantum}\ }\textbf {\bibinfo {volume} {3}},\ \bibinfo {pages} {030325}
  (\bibinfo {year} {2022})}\BibitemShut {NoStop}%
\bibitem [{\citenamefont {Henry}\ \emph {et~al.}(2021)\citenamefont {Henry},
  \citenamefont {Thabet}, \citenamefont {Dalyac},\ and\ \citenamefont
  {Henriet}}]{Henry_2021}%
  \BibitemOpen
  \bibfield  {author} {\bibinfo {author} {\bibfnamefont {L.-P.}\ \bibnamefont
  {Henry}}, \bibinfo {author} {\bibfnamefont {S.}~\bibnamefont {Thabet}},
  \bibinfo {author} {\bibfnamefont {C.}~\bibnamefont {Dalyac}}, \ and\ \bibinfo
  {author} {\bibfnamefont {L.}~\bibnamefont {Henriet}},\ }\href {\doibase
  10.1103/physreva.104.032416} {\bibfield  {journal} {\bibinfo  {journal}
  {Physical Review A}\ }\textbf {\bibinfo {volume} {104}} (\bibinfo {year}
  {2021}),\ 10.1103/physreva.104.032416}\BibitemShut {NoStop}%
\bibitem [{\citenamefont {Leclerc}\ \emph {et~al.}(2022)\citenamefont
  {Leclerc}, \citenamefont {Ortiz-Guitierrez}, \citenamefont {Grijalva},
  \citenamefont {Albrecht}, \citenamefont {Cline}, \citenamefont {Elfving},
  \citenamefont {Signoles}, \citenamefont {Henriet}, \citenamefont {Bimbo},
  \citenamefont {Sheikh}, \citenamefont {Shah}, \citenamefont {Andrea},
  \citenamefont {Ishtiaq}, \citenamefont {Duarte}, \citenamefont {Mugel} \emph
  {et~al.}}]{leclerc2022financial}%
  \BibitemOpen
  \bibfield  {author} {\bibinfo {author} {\bibfnamefont {L.}~\bibnamefont
  {Leclerc}}, \bibinfo {author} {\bibfnamefont {L.}~\bibnamefont
  {Ortiz-Guitierrez}}, \bibinfo {author} {\bibfnamefont {S.}~\bibnamefont
  {Grijalva}}, \bibinfo {author} {\bibfnamefont {B.}~\bibnamefont {Albrecht}},
  \bibinfo {author} {\bibfnamefont {J.~R.~K.}\ \bibnamefont {Cline}}, \bibinfo
  {author} {\bibfnamefont {V.~E.}\ \bibnamefont {Elfving}}, \bibinfo {author}
  {\bibfnamefont {A.}~\bibnamefont {Signoles}}, \bibinfo {author}
  {\bibfnamefont {L.}~\bibnamefont {Henriet}}, \bibinfo {author} {\bibfnamefont
  {G.~D.}\ \bibnamefont {Bimbo}}, \bibinfo {author} {\bibfnamefont {U.~A.}\
  \bibnamefont {Sheikh}}, \bibinfo {author} {\bibfnamefont {M.}~\bibnamefont
  {Shah}}, \bibinfo {author} {\bibfnamefont {L.}~\bibnamefont {Andrea}},
  \bibinfo {author} {\bibfnamefont {F.}~\bibnamefont {Ishtiaq}}, \bibinfo
  {author} {\bibfnamefont {A.}~\bibnamefont {Duarte}}, \bibinfo {author}
  {\bibfnamefont {S.}~\bibnamefont {Mugel}},  \emph {et~al.},\ }\href@noop {}
  {\  (\bibinfo {year} {2022})},\ \Eprint {http://arxiv.org/abs/2212.03223}
  {arXiv:2212.03223 [quant-ph]} \BibitemShut {NoStop}%
\bibitem [{\citenamefont {Albrecht}\ \emph {et~al.}(2023)\citenamefont
  {Albrecht}, \citenamefont {Dalyac}, \citenamefont {Leclerc}, \citenamefont
  {Ortiz-Guti\'errez}, \citenamefont {Thabet}, \citenamefont {D'Arcangelo},
  \citenamefont {Cline}, \citenamefont {Elfving}, \citenamefont
  {Lassabli\`ere}, \citenamefont {Silv\'erio}, \citenamefont {Ximenez},
  \citenamefont {Henry}, \citenamefont {Signoles},\ and\ \citenamefont
  {Henriet}}]{PhysRevA.107.042615}%
  \BibitemOpen
  \bibfield  {author} {\bibinfo {author} {\bibfnamefont {B.}~\bibnamefont
  {Albrecht}}, \bibinfo {author} {\bibfnamefont {C.}~\bibnamefont {Dalyac}},
  \bibinfo {author} {\bibfnamefont {L.}~\bibnamefont {Leclerc}}, \bibinfo
  {author} {\bibfnamefont {L.}~\bibnamefont {Ortiz-Guti\'errez}}, \bibinfo
  {author} {\bibfnamefont {S.}~\bibnamefont {Thabet}}, \bibinfo {author}
  {\bibfnamefont {M.}~\bibnamefont {D'Arcangelo}}, \bibinfo {author}
  {\bibfnamefont {J.~R.~K.}\ \bibnamefont {Cline}}, \bibinfo {author}
  {\bibfnamefont {V.~E.}\ \bibnamefont {Elfving}}, \bibinfo {author}
  {\bibfnamefont {L.}~\bibnamefont {Lassabli\`ere}}, \bibinfo {author}
  {\bibfnamefont {H.}~\bibnamefont {Silv\'erio}}, \bibinfo {author}
  {\bibfnamefont {B.}~\bibnamefont {Ximenez}}, \bibinfo {author} {\bibfnamefont
  {L.-P.}\ \bibnamefont {Henry}}, \bibinfo {author} {\bibfnamefont
  {A.}~\bibnamefont {Signoles}}, \ and\ \bibinfo {author} {\bibfnamefont
  {L.}~\bibnamefont {Henriet}},\ }\href {\doibase 10.1103/PhysRevA.107.042615}
  {\bibfield  {journal} {\bibinfo  {journal} {Phys. Rev. A}\ }\textbf {\bibinfo
  {volume} {107}},\ \bibinfo {pages} {042615} (\bibinfo {year}
  {2023})}\BibitemShut {NoStop}%
\bibitem [{\citenamefont {Michel}\ \emph {et~al.}(2023)\citenamefont {Michel},
  \citenamefont {Grijalva}, \citenamefont {Henriet}, \citenamefont {Domain},\
  and\ \citenamefont {Browaeys}}]{PhysRevA.107.042602}%
  \BibitemOpen
  \bibfield  {author} {\bibinfo {author} {\bibfnamefont {A.}~\bibnamefont
  {Michel}}, \bibinfo {author} {\bibfnamefont {S.}~\bibnamefont {Grijalva}},
  \bibinfo {author} {\bibfnamefont {L.}~\bibnamefont {Henriet}}, \bibinfo
  {author} {\bibfnamefont {C.}~\bibnamefont {Domain}}, \ and\ \bibinfo {author}
  {\bibfnamefont {A.}~\bibnamefont {Browaeys}},\ }\href {\doibase
  10.1103/PhysRevA.107.042602} {\bibfield  {journal} {\bibinfo  {journal}
  {Phys. Rev. A}\ }\textbf {\bibinfo {volume} {107}},\ \bibinfo {pages}
  {042602} (\bibinfo {year} {2023})}\BibitemShut {NoStop}%
\bibitem [{\citenamefont {D'Arcangelo}\ \emph {et~al.}(2023)\citenamefont
  {D'Arcangelo}, \citenamefont {Loco}, \citenamefont {team}, \citenamefont
  {Gouraud}, \citenamefont {Angebault}, \citenamefont {Sueiro}, \citenamefont
  {Monmarché}, \citenamefont {Forêt}, \citenamefont {Henry}, \citenamefont
  {Henriet},\ and\ \citenamefont {Piquemal}}]{darcangelo2023leveraging}%
  \BibitemOpen
  \bibfield  {author} {\bibinfo {author} {\bibfnamefont {M.}~\bibnamefont
  {D'Arcangelo}}, \bibinfo {author} {\bibfnamefont {D.}~\bibnamefont {Loco}},
  \bibinfo {author} {\bibfnamefont {F.}~\bibnamefont {team}}, \bibinfo {author}
  {\bibfnamefont {N.}~\bibnamefont {Gouraud}}, \bibinfo {author} {\bibfnamefont
  {S.}~\bibnamefont {Angebault}}, \bibinfo {author} {\bibfnamefont
  {J.}~\bibnamefont {Sueiro}}, \bibinfo {author} {\bibfnamefont
  {P.}~\bibnamefont {Monmarché}}, \bibinfo {author} {\bibfnamefont
  {J.}~\bibnamefont {Forêt}}, \bibinfo {author} {\bibfnamefont {L.-P.}\
  \bibnamefont {Henry}}, \bibinfo {author} {\bibfnamefont {L.}~\bibnamefont
  {Henriet}}, \ and\ \bibinfo {author} {\bibfnamefont {J.-P.}\ \bibnamefont
  {Piquemal}},\ }\href@noop {} {\  (\bibinfo {year} {2023})},\ \Eprint
  {http://arxiv.org/abs/2309.12129} {arXiv:2309.12129 [quant-ph]} \BibitemShut
  {NoStop}%
\bibitem [{\citenamefont {Urban}\ \emph {et~al.}(2009)\citenamefont {Urban},
  \citenamefont {Johnson}, \citenamefont {Henage}, \citenamefont {Isenhower},
  \citenamefont {Yavuz}, \citenamefont {Walker},\ and\ \citenamefont
  {Saffman}}]{Urban_2009}%
  \BibitemOpen
  \bibfield  {author} {\bibinfo {author} {\bibfnamefont {E.}~\bibnamefont
  {Urban}}, \bibinfo {author} {\bibfnamefont {T.~A.}\ \bibnamefont {Johnson}},
  \bibinfo {author} {\bibfnamefont {T.}~\bibnamefont {Henage}}, \bibinfo
  {author} {\bibfnamefont {L.}~\bibnamefont {Isenhower}}, \bibinfo {author}
  {\bibfnamefont {D.~D.}\ \bibnamefont {Yavuz}}, \bibinfo {author}
  {\bibfnamefont {T.~G.}\ \bibnamefont {Walker}}, \ and\ \bibinfo {author}
  {\bibfnamefont {M.}~\bibnamefont {Saffman}},\ }\href {\doibase
  10.1038/nphys1178} {\bibfield  {journal} {\bibinfo  {journal} {Nature
  Physics}\ }\textbf {\bibinfo {volume} {5}},\ \bibinfo {pages} {110} (\bibinfo
  {year} {2009})}\BibitemShut {NoStop}%
\bibitem [{\citenamefont {Wurtz}\ \emph {et~al.}(2023)\citenamefont {Wurtz},
  \citenamefont {Bylinskii}, \citenamefont {Braverman}, \citenamefont
  {Amato-Grill}, \citenamefont {Cantu}, \citenamefont {Huber}, \citenamefont
  {Lukin}, \citenamefont {Liu}, \citenamefont {Weinberg}, \citenamefont {Long},
  \citenamefont {Wang}, \citenamefont {Gemelke},\ and\ \citenamefont
  {Keesling}}]{wurtz2023aquila}%
  \BibitemOpen
  \bibfield  {author} {\bibinfo {author} {\bibfnamefont {J.}~\bibnamefont
  {Wurtz}}, \bibinfo {author} {\bibfnamefont {A.}~\bibnamefont {Bylinskii}},
  \bibinfo {author} {\bibfnamefont {B.}~\bibnamefont {Braverman}}, \bibinfo
  {author} {\bibfnamefont {J.}~\bibnamefont {Amato-Grill}}, \bibinfo {author}
  {\bibfnamefont {S.~H.}\ \bibnamefont {Cantu}}, \bibinfo {author}
  {\bibfnamefont {F.}~\bibnamefont {Huber}}, \bibinfo {author} {\bibfnamefont
  {A.}~\bibnamefont {Lukin}}, \bibinfo {author} {\bibfnamefont
  {F.}~\bibnamefont {Liu}}, \bibinfo {author} {\bibfnamefont {P.}~\bibnamefont
  {Weinberg}}, \bibinfo {author} {\bibfnamefont {J.}~\bibnamefont {Long}},
  \bibinfo {author} {\bibfnamefont {S.-T.}\ \bibnamefont {Wang}}, \bibinfo
  {author} {\bibfnamefont {N.}~\bibnamefont {Gemelke}}, \ and\ \bibinfo
  {author} {\bibfnamefont {A.}~\bibnamefont {Keesling}},\ }\href@noop {} {\
  (\bibinfo {year} {2023})},\ \Eprint {http://arxiv.org/abs/2306.11727}
  {arXiv:2306.11727 [quant-ph]} \BibitemShut {NoStop}%
\bibitem [{\citenamefont {Vandersypen}\ and\ \citenamefont
  {Chuang}(2005)}]{Vandersypen05}%
  \BibitemOpen
  \bibfield  {author} {\bibinfo {author} {\bibfnamefont {L.~M.~K.}\
  \bibnamefont {Vandersypen}}\ and\ \bibinfo {author} {\bibfnamefont {I.~L.}\
  \bibnamefont {Chuang}},\ }\href {\doibase 10.1103/RevModPhys.76.1037}
  {\bibfield  {journal} {\bibinfo  {journal} {Rev. Mod. Phys.}\ }\textbf
  {\bibinfo {volume} {76}},\ \bibinfo {pages} {1037} (\bibinfo {year}
  {2005})}\BibitemShut {NoStop}%
\bibitem [{\citenamefont {Landsman}\ \emph {et~al.}(2019)\citenamefont
  {Landsman}, \citenamefont {Wu}, \citenamefont {Leung}, \citenamefont {Zhu},
  \citenamefont {Linke}, \citenamefont {Brown}, \citenamefont {Duan},\ and\
  \citenamefont {Monroe}}]{PhysRevA.100.022332}%
  \BibitemOpen
  \bibfield  {author} {\bibinfo {author} {\bibfnamefont {K.~A.}\ \bibnamefont
  {Landsman}}, \bibinfo {author} {\bibfnamefont {Y.}~\bibnamefont {Wu}},
  \bibinfo {author} {\bibfnamefont {P.~H.}\ \bibnamefont {Leung}}, \bibinfo
  {author} {\bibfnamefont {D.}~\bibnamefont {Zhu}}, \bibinfo {author}
  {\bibfnamefont {N.~M.}\ \bibnamefont {Linke}}, \bibinfo {author}
  {\bibfnamefont {K.~R.}\ \bibnamefont {Brown}}, \bibinfo {author}
  {\bibfnamefont {L.}~\bibnamefont {Duan}}, \ and\ \bibinfo {author}
  {\bibfnamefont {C.}~\bibnamefont {Monroe}},\ }\href {\doibase
  10.1103/PhysRevA.100.022332} {\bibfield  {journal} {\bibinfo  {journal}
  {Phys. Rev. A}\ }\textbf {\bibinfo {volume} {100}},\ \bibinfo {pages}
  {022332} (\bibinfo {year} {2019})}\BibitemShut {NoStop}%
\bibitem [{\citenamefont {Bartolucci}\ \emph {et~al.}(2021)\citenamefont
  {Bartolucci}, \citenamefont {Birchall}, \citenamefont {Bombin}, \citenamefont
  {Cable}, \citenamefont {Dawson}, \citenamefont {Gimeno-Segovia},
  \citenamefont {Johnston}, \citenamefont {Kieling}, \citenamefont {Nickerson},
  \citenamefont {Pant}, \citenamefont {Pastawski}, \citenamefont {Rudolph},\
  and\ \citenamefont {Sparrow}}]{bartolucci2021fusionbased}%
  \BibitemOpen
  \bibfield  {author} {\bibinfo {author} {\bibfnamefont {S.}~\bibnamefont
  {Bartolucci}}, \bibinfo {author} {\bibfnamefont {P.}~\bibnamefont
  {Birchall}}, \bibinfo {author} {\bibfnamefont {H.}~\bibnamefont {Bombin}},
  \bibinfo {author} {\bibfnamefont {H.}~\bibnamefont {Cable}}, \bibinfo
  {author} {\bibfnamefont {C.}~\bibnamefont {Dawson}}, \bibinfo {author}
  {\bibfnamefont {M.}~\bibnamefont {Gimeno-Segovia}}, \bibinfo {author}
  {\bibfnamefont {E.}~\bibnamefont {Johnston}}, \bibinfo {author}
  {\bibfnamefont {K.}~\bibnamefont {Kieling}}, \bibinfo {author} {\bibfnamefont
  {N.}~\bibnamefont {Nickerson}}, \bibinfo {author} {\bibfnamefont
  {M.}~\bibnamefont {Pant}}, \bibinfo {author} {\bibfnamefont {F.}~\bibnamefont
  {Pastawski}}, \bibinfo {author} {\bibfnamefont {T.}~\bibnamefont {Rudolph}},
  \ and\ \bibinfo {author} {\bibfnamefont {C.}~\bibnamefont {Sparrow}},\
  }\href@noop {} {\  (\bibinfo {year} {2021})},\ \Eprint
  {http://arxiv.org/abs/2101.09310} {arXiv:2101.09310 [quant-ph]} \BibitemShut
  {NoStop}%
\bibitem [{\citenamefont {Bluvstein}\ \emph {et~al.}(2022)\citenamefont
  {Bluvstein}, \citenamefont {Levine}, \citenamefont {Semeghini}, \citenamefont
  {Wang}, \citenamefont {Ebadi}, \citenamefont {Kalinowski}, \citenamefont
  {Keesling}, \citenamefont {Maskara}, \citenamefont {Pichler}, \citenamefont
  {Greiner}, \citenamefont {Vuleti{\'{c}}},\ and\ \citenamefont
  {Lukin}}]{Bluvstein_2022}%
  \BibitemOpen
  \bibfield  {author} {\bibinfo {author} {\bibfnamefont {D.}~\bibnamefont
  {Bluvstein}}, \bibinfo {author} {\bibfnamefont {H.}~\bibnamefont {Levine}},
  \bibinfo {author} {\bibfnamefont {G.}~\bibnamefont {Semeghini}}, \bibinfo
  {author} {\bibfnamefont {T.~T.}\ \bibnamefont {Wang}}, \bibinfo {author}
  {\bibfnamefont {S.}~\bibnamefont {Ebadi}}, \bibinfo {author} {\bibfnamefont
  {M.}~\bibnamefont {Kalinowski}}, \bibinfo {author} {\bibfnamefont
  {A.}~\bibnamefont {Keesling}}, \bibinfo {author} {\bibfnamefont
  {N.}~\bibnamefont {Maskara}}, \bibinfo {author} {\bibfnamefont
  {H.}~\bibnamefont {Pichler}}, \bibinfo {author} {\bibfnamefont
  {M.}~\bibnamefont {Greiner}}, \bibinfo {author} {\bibfnamefont
  {V.}~\bibnamefont {Vuleti{\'{c}}}}, \ and\ \bibinfo {author} {\bibfnamefont
  {M.~D.}\ \bibnamefont {Lukin}},\ }\href {\doibase 10.1038/s41586-022-04592-6}
  {\bibfield  {journal} {\bibinfo  {journal} {Nature}\ }\textbf {\bibinfo
  {volume} {604}},\ \bibinfo {pages} {451} (\bibinfo {year}
  {2022})}\BibitemShut {NoStop}%
\bibitem [{\citenamefont {Pan}\ and\ \citenamefont
  {Zhang}(2021)}]{pan2021simulating}%
  \BibitemOpen
  \bibfield  {author} {\bibinfo {author} {\bibfnamefont {F.}~\bibnamefont
  {Pan}}\ and\ \bibinfo {author} {\bibfnamefont {P.}~\bibnamefont {Zhang}},\
  }\href@noop {} {\  (\bibinfo {year} {2021})},\ \Eprint
  {http://arxiv.org/abs/2103.03074} {arXiv:2103.03074 [quant-ph]} \BibitemShut
  {NoStop}%
\bibitem [{\citenamefont {D{\'{\i} }ez-Valle}\ \emph
  {et~al.}(2023)\citenamefont {D{\'{\i} }ez-Valle}, \citenamefont {Porras},\
  and\ \citenamefont {Garc{\'{\i}}a-Ripoll}}]{D_ez_Valle_2023}%
  \BibitemOpen
  \bibfield  {author} {\bibinfo {author} {\bibfnamefont {P.}~\bibnamefont
  {D{\'{\i} }ez-Valle}}, \bibinfo {author} {\bibfnamefont {D.}~\bibnamefont
  {Porras}}, \ and\ \bibinfo {author} {\bibfnamefont {J.~J.}\ \bibnamefont
  {Garc{\'{\i}}a-Ripoll}},\ }\href {\doibase 10.1103/physrevlett.130.050601}
  {\bibfield  {journal} {\bibinfo  {journal} {Physical Review Letters}\
  }\textbf {\bibinfo {volume} {130}} (\bibinfo {year} {2023}),\
  10.1103/physrevlett.130.050601}\BibitemShut {NoStop}%
\bibitem [{\citenamefont {Lucas}(2014)}]{Lucas_2014}%
  \BibitemOpen
  \bibfield  {author} {\bibinfo {author} {\bibfnamefont {A.}~\bibnamefont
  {Lucas}},\ }\href {\doibase 10.3389/fphy.2014.00005} {\bibfield  {journal}
  {\bibinfo  {journal} {Frontiers in Physics}\ }\textbf {\bibinfo {volume} {2}}
  (\bibinfo {year} {2014}),\ 10.3389/fphy.2014.00005}\BibitemShut {NoStop}%
\bibitem [{\citenamefont {O'Gorman}\ \emph {et~al.}(2019)\citenamefont
  {O'Gorman}, \citenamefont {Huggins}, \citenamefont {Rieffel},\ and\
  \citenamefont {Whaley}}]{ogorman2019generalized}%
  \BibitemOpen
  \bibfield  {author} {\bibinfo {author} {\bibfnamefont {B.}~\bibnamefont
  {O'Gorman}}, \bibinfo {author} {\bibfnamefont {W.~J.}\ \bibnamefont
  {Huggins}}, \bibinfo {author} {\bibfnamefont {E.~G.}\ \bibnamefont
  {Rieffel}}, \ and\ \bibinfo {author} {\bibfnamefont {K.~B.}\ \bibnamefont
  {Whaley}},\ }\href@noop {} {\  (\bibinfo {year} {2019})},\ \Eprint
  {http://arxiv.org/abs/1905.05118} {arXiv:1905.05118 [quant-ph]} \BibitemShut
  {NoStop}%
\bibitem [{\citenamefont {Elfving}\ \emph {et~al.}(2021)\citenamefont
  {Elfving}, \citenamefont {Millaruelo}, \citenamefont {G\'amez},\ and\
  \citenamefont {Gogolin}}]{Elfving2021}%
  \BibitemOpen
  \bibfield  {author} {\bibinfo {author} {\bibfnamefont {V.~E.}\ \bibnamefont
  {Elfving}}, \bibinfo {author} {\bibfnamefont {M.}~\bibnamefont {Millaruelo}},
  \bibinfo {author} {\bibfnamefont {J.~A.}\ \bibnamefont {G\'amez}}, \ and\
  \bibinfo {author} {\bibfnamefont {C.}~\bibnamefont {Gogolin}},\ }\href
  {\doibase 10.1103/PhysRevA.103.032605} {\bibfield  {journal} {\bibinfo
  {journal} {Phys. Rev. A}\ }\textbf {\bibinfo {volume} {103}},\ \bibinfo
  {pages} {032605} (\bibinfo {year} {2021})}\BibitemShut {NoStop}%
\bibitem [{\citenamefont {Seitz}\ \emph {et~al.}(2024)\citenamefont {Seitz},
  \citenamefont {Heim}, \citenamefont {Moutinho}, \citenamefont {Guichard},
  \citenamefont {Abramavicius}, \citenamefont {Wennersteen}, \citenamefont
  {Both}, \citenamefont {Quelle}, \citenamefont {de~Groot}, \citenamefont
  {Velikova}, \citenamefont {Elfving},\ and\ \citenamefont
  {Dagrada}}]{seitz2024qadence}%
  \BibitemOpen
  \bibfield  {author} {\bibinfo {author} {\bibfnamefont {D.}~\bibnamefont
  {Seitz}}, \bibinfo {author} {\bibfnamefont {N.}~\bibnamefont {Heim}},
  \bibinfo {author} {\bibfnamefont {J.~P.}\ \bibnamefont {Moutinho}}, \bibinfo
  {author} {\bibfnamefont {R.}~\bibnamefont {Guichard}}, \bibinfo {author}
  {\bibfnamefont {V.}~\bibnamefont {Abramavicius}}, \bibinfo {author}
  {\bibfnamefont {A.}~\bibnamefont {Wennersteen}}, \bibinfo {author}
  {\bibfnamefont {G.-J.}\ \bibnamefont {Both}}, \bibinfo {author}
  {\bibfnamefont {A.}~\bibnamefont {Quelle}}, \bibinfo {author} {\bibfnamefont
  {C.}~\bibnamefont {de~Groot}}, \bibinfo {author} {\bibfnamefont {G.~V.}\
  \bibnamefont {Velikova}}, \bibinfo {author} {\bibfnamefont {V.~E.}\
  \bibnamefont {Elfving}}, \ and\ \bibinfo {author} {\bibfnamefont
  {M.}~\bibnamefont {Dagrada}},\ }\href@noop {} {\  (\bibinfo {year} {2024})},\
  \Eprint {http://arxiv.org/abs/2401.09915} {arXiv:2401.09915 [quant-ph]}
  \BibitemShut {NoStop}%
\bibitem [{\citenamefont {O'Brien}\ \emph {et~al.}(2022)\citenamefont
  {O'Brien}, \citenamefont {Anselmetti}, \citenamefont {Gkritsis},
  \citenamefont {Elfving}, \citenamefont {Polla}, \citenamefont {Huggins},
  \citenamefont {Oumarou}, \citenamefont {Kechedzhi}, \citenamefont {Abanin},
  \citenamefont {Acharya}, \citenamefont {Aleiner}, \citenamefont {Allen},
  \citenamefont {Andersen}, \citenamefont {Anderson}, \citenamefont {Ansmann}
  \emph {et~al.}}]{obrien2022purificationbased}%
  \BibitemOpen
  \bibfield  {author} {\bibinfo {author} {\bibfnamefont {T.~E.}\ \bibnamefont
  {O'Brien}}, \bibinfo {author} {\bibfnamefont {G.}~\bibnamefont {Anselmetti}},
  \bibinfo {author} {\bibfnamefont {F.}~\bibnamefont {Gkritsis}}, \bibinfo
  {author} {\bibfnamefont {V.~E.}\ \bibnamefont {Elfving}}, \bibinfo {author}
  {\bibfnamefont {S.}~\bibnamefont {Polla}}, \bibinfo {author} {\bibfnamefont
  {W.~J.}\ \bibnamefont {Huggins}}, \bibinfo {author} {\bibfnamefont
  {O.}~\bibnamefont {Oumarou}}, \bibinfo {author} {\bibfnamefont
  {K.}~\bibnamefont {Kechedzhi}}, \bibinfo {author} {\bibfnamefont
  {D.}~\bibnamefont {Abanin}}, \bibinfo {author} {\bibfnamefont
  {R.}~\bibnamefont {Acharya}}, \bibinfo {author} {\bibfnamefont
  {I.}~\bibnamefont {Aleiner}}, \bibinfo {author} {\bibfnamefont
  {R.}~\bibnamefont {Allen}}, \bibinfo {author} {\bibfnamefont {T.~I.}\
  \bibnamefont {Andersen}}, \bibinfo {author} {\bibfnamefont {K.}~\bibnamefont
  {Anderson}}, \bibinfo {author} {\bibfnamefont {M.}~\bibnamefont {Ansmann}},
  \emph {et~al.},\ }\href@noop {} {\  (\bibinfo {year} {2022})},\ \Eprint
  {http://arxiv.org/abs/2210.10799} {arXiv:2210.10799 [quant-ph]} \BibitemShut
  {NoStop}%
\bibitem [{\citenamefont {Jensen}(2017)}]{jensen2017introduction}%
  \BibitemOpen
  \bibfield  {author} {\bibinfo {author} {\bibfnamefont {F.}~\bibnamefont
  {Jensen}},\ }\href@noop {} {\emph {\bibinfo {title} {Introduction to
  computational chemistry}}}\ (\bibinfo  {publisher} {John wiley \& sons},\
  \bibinfo {year} {2017})\BibitemShut {NoStop}%
\bibitem [{\citenamefont {Geen}\ and\ \citenamefont {Freeman}(1991)}]{Geen91}%
  \BibitemOpen
  \bibfield  {author} {\bibinfo {author} {\bibfnamefont {H.}~\bibnamefont
  {Geen}}\ and\ \bibinfo {author} {\bibfnamefont {R.}~\bibnamefont {Freeman}},\
  }\href {https://doi.org/10.1016/0022-2364(91)90034-Q} {\bibfield  {journal}
  {\bibinfo  {journal} {Journal of Magnetic Resonance (1969)}\ }\textbf
  {\bibinfo {volume} {93}},\ \bibinfo {pages} {93} (\bibinfo {year}
  {1991})}\BibitemShut {NoStop}%
\bibitem [{\citenamefont {Yang}\ \emph {et~al.}(2017)\citenamefont {Yang},
  \citenamefont {Rahmani}, \citenamefont {Shabani}, \citenamefont {Neven},\
  and\ \citenamefont {Chamon}}]{PhysRevX.7.021027}%
  \BibitemOpen
  \bibfield  {author} {\bibinfo {author} {\bibfnamefont {Z.-C.}\ \bibnamefont
  {Yang}}, \bibinfo {author} {\bibfnamefont {A.}~\bibnamefont {Rahmani}},
  \bibinfo {author} {\bibfnamefont {A.}~\bibnamefont {Shabani}}, \bibinfo
  {author} {\bibfnamefont {H.}~\bibnamefont {Neven}}, \ and\ \bibinfo {author}
  {\bibfnamefont {C.}~\bibnamefont {Chamon}},\ }\href {\doibase
  10.1103/PhysRevX.7.021027} {\bibfield  {journal} {\bibinfo  {journal} {Phys.
  Rev. X}\ }\textbf {\bibinfo {volume} {7}},\ \bibinfo {pages} {021027}
  (\bibinfo {year} {2017})}\BibitemShut {NoStop}%
\bibitem [{\citenamefont {Caneva}\ \emph {et~al.}(2011)\citenamefont {Caneva},
  \citenamefont {Calarco},\ and\ \citenamefont {Montangero}}]{Caneva11}%
  \BibitemOpen
  \bibfield  {author} {\bibinfo {author} {\bibfnamefont {T.}~\bibnamefont
  {Caneva}}, \bibinfo {author} {\bibfnamefont {T.}~\bibnamefont {Calarco}}, \
  and\ \bibinfo {author} {\bibfnamefont {S.}~\bibnamefont {Montangero}},\
  }\href {\doibase 10.1103/PhysRevA.84.022326} {\bibfield  {journal} {\bibinfo
  {journal} {Phys. Rev. A}\ }\textbf {\bibinfo {volume} {84}},\ \bibinfo
  {pages} {022326} (\bibinfo {year} {2011})}\BibitemShut {NoStop}%
\bibitem [{\citenamefont {Khaneja}\ \emph {et~al.}(2005)\citenamefont
  {Khaneja}, \citenamefont {Reiss}, \citenamefont {Kehlet}, \citenamefont
  {Schulte-Herbr{\"u}ggen},\ and\ \citenamefont {Glaser}}]{Khaneja05}%
  \BibitemOpen
  \bibfield  {author} {\bibinfo {author} {\bibfnamefont {N.}~\bibnamefont
  {Khaneja}}, \bibinfo {author} {\bibfnamefont {T.}~\bibnamefont {Reiss}},
  \bibinfo {author} {\bibfnamefont {C.}~\bibnamefont {Kehlet}}, \bibinfo
  {author} {\bibfnamefont {T.}~\bibnamefont {Schulte-Herbr{\"u}ggen}}, \ and\
  \bibinfo {author} {\bibfnamefont {S.~J.}\ \bibnamefont {Glaser}},\ }\href
  {https://doi.org/10.1016/j.jmr.2004.11.004} {\bibfield  {journal} {\bibinfo
  {journal} {Journal of Magnetic Resonance}\ }\textbf {\bibinfo {volume}
  {172}},\ \bibinfo {pages} {296} (\bibinfo {year} {2005})}\BibitemShut
  {NoStop}%
\bibitem [{\citenamefont {Leng}\ \emph {et~al.}(2022)\citenamefont {Leng},
  \citenamefont {Peng}, \citenamefont {Qiao}, \citenamefont {Lin},\ and\
  \citenamefont {Wu}}]{Leng22}%
  \BibitemOpen
  \bibfield  {author} {\bibinfo {author} {\bibfnamefont {J.}~\bibnamefont
  {Leng}}, \bibinfo {author} {\bibfnamefont {Y.}~\bibnamefont {Peng}}, \bibinfo
  {author} {\bibfnamefont {Y.-L.}\ \bibnamefont {Qiao}}, \bibinfo {author}
  {\bibfnamefont {M.}~\bibnamefont {Lin}}, \ and\ \bibinfo {author}
  {\bibfnamefont {X.}~\bibnamefont {Wu}},\ }\bibfield  {booktitle} {\emph
  {\bibinfo {booktitle} {Advances in Neural Information Processing Systems}},\
  }\href
  {https://proceedings.neurips.cc/paper_files/paper/2022/file/1e70ac91ad26ba5b24cf11b12a1f90fe-Paper-Conference.pdf}
  {\ \textbf {\bibinfo {volume} {35}},\ \bibinfo {pages} {4707} (\bibinfo
  {year} {2022})}\BibitemShut {NoStop}%
\bibitem [{\citenamefont {Shahriari}\ \emph {et~al.}(2016)\citenamefont
  {Shahriari}, \citenamefont {Swersky}, \citenamefont {Wang}, \citenamefont
  {Adams},\ and\ \citenamefont {de~Freitas}}]{Shahriari15}%
  \BibitemOpen
  \bibfield  {author} {\bibinfo {author} {\bibfnamefont {B.}~\bibnamefont
  {Shahriari}}, \bibinfo {author} {\bibfnamefont {K.}~\bibnamefont {Swersky}},
  \bibinfo {author} {\bibfnamefont {Z.}~\bibnamefont {Wang}}, \bibinfo {author}
  {\bibfnamefont {R.~P.}\ \bibnamefont {Adams}}, \ and\ \bibinfo {author}
  {\bibfnamefont {N.}~\bibnamefont {de~Freitas}},\ }\href {\doibase
  10.1109/JPROC.2015.2494218} {\bibfield  {journal} {\bibinfo  {journal}
  {Proceedings of the IEEE}\ }\textbf {\bibinfo {volume} {104}},\ \bibinfo
  {pages} {148} (\bibinfo {year} {2016})}\BibitemShut {NoStop}%
\bibitem [{\citenamefont {Frazier}(2018)}]{Frazier18}%
  \BibitemOpen
  \bibfield  {author} {\bibinfo {author} {\bibfnamefont {P.~I.}\ \bibnamefont
  {Frazier}},\ }in\ \href {https://doi.org/10.1287/educ.2018.0188} {\emph
  {\bibinfo {booktitle} {Recent advances in optimization and modeling of
  contemporary problems}}}\ (\bibinfo  {publisher} {Informs},\ \bibinfo {year}
  {2018})\ pp.\ \bibinfo {pages} {255--278}\BibitemShut {NoStop}%
\bibitem [{\citenamefont {Notarnicola}\ \emph {et~al.}(2023)\citenamefont
  {Notarnicola}, \citenamefont {Elben}, \citenamefont {Lahaye}, \citenamefont
  {Browaeys}, \citenamefont {Montangero},\ and\ \citenamefont
  {Vermersch}}]{notarnicola2023randomized}%
  \BibitemOpen
  \bibfield  {author} {\bibinfo {author} {\bibfnamefont {S.}~\bibnamefont
  {Notarnicola}}, \bibinfo {author} {\bibfnamefont {A.}~\bibnamefont {Elben}},
  \bibinfo {author} {\bibfnamefont {T.}~\bibnamefont {Lahaye}}, \bibinfo
  {author} {\bibfnamefont {A.}~\bibnamefont {Browaeys}}, \bibinfo {author}
  {\bibfnamefont {S.}~\bibnamefont {Montangero}}, \ and\ \bibinfo {author}
  {\bibfnamefont {B.}~\bibnamefont {Vermersch}},\ }\href@noop {} {\  (\bibinfo
  {year} {2023})},\ \Eprint {http://arxiv.org/abs/2112.11046} {arXiv:2112.11046
  [quant-ph]} \BibitemShut {NoStop}%
\end{thebibliography}%

\end{document}